\documentclass[twocolumn,prb,aps,amsmath,showpacs,citeautoscript]{revtex4}

\def\PRB{Phys. Rev. B}
\usepackage{graphicx}
\usepackage{natbib}
\begin{document}
\title{Phonon Spectra and Thermal Properties of Some fcc Metals Using  
the Embedded-Atom Method}      
\author{Q. Bian$^{\dagger}$}
\email{bianq2@univmail.cis.mcmaster.ca}    
\author{S.K. Bose}      
\author{R.C. Shukla}  

\affiliation{Department of Physics, Brock University, St.\ Catharines, 
Ontario
L2S 3A1, Canada}
\date{\today}
\begin{abstract} 
By employing the analytic embedded-atom potentials of Mei {\it et al.} 
[Phys. Rev. B 43, 4653 (1991)] we have calculated the phonon dispersion 
spectra for six fcc metals: Cu, Ag, Au, Ni, Pd and Pt. We have also 
investigated thermal properties of these metals within the quasiharmonic 
approximation. Results for the lattice constants, coefficients of linear 
thermal expansion, isothermal and adiabatic bulk moduli, heat capacities 
at constant volume and constant pressure, Debye temperatures
 and Gr\"uneisen parameters as a function of temperature are presented. 
 The computed results are compared with the available experimental data. 
 The comparison shows a generally good agreement between the calculated 
 and experimental values for all thermodynamic properties studied. Isothermal 
 and adiabatic bulk moduli and the specific heats are reproduced reasonably 
 well, while  the Gr\"uneisen parameter and Debye temperature are underestimated 
 by about 10\%. The calculated phonon frequencies for Ag and Cu agree well 
 with the results from inelastic neutron scattering experiments. However, 
 there is considerable room for improvement in the phonon frequencies
for Ni, Pd, Pt and Au, particularly at high phonon wave vectors close 
to the  Brillouin zone boundary. The coefficient of linear thermal expansion 
is underestimated in most cases except for Pt and  Au. The results 
are good for Pt up to 1000K and for Au up to 500 K.
\end{abstract} 

\pacs{63.20.-e, 65.40.-b}

\maketitle 
\section {INTRODUCTION} 

The inadequacy of pure pair-potential models to describe metallic cohesion is
well-known and has been adequately documented\cite{pair-pot,carlsson}. 
Various approaches, at varying
 levels of sophistication, have been used to address the volume-dependence of the
 energy of a metallic system originating from the presence of the interacting electron
 gas. These range from using volume-dependent parameters in the pair potential
 itself\cite{shukla} to writing the total energy as a sum of pair potentials plus
 an empirical volume/density-dependent 
 term\cite{Finnis26,Legrand,Rosato,Cleri,Zhang} or an electronic band(bond) energy 
term\cite{Pettifor86,Pettifor87}. The latter is often written in terms of the
first few moments of the electronic density of states in the tight-binding 
approximation\cite{Ducastelle70,Ducastelle71}. For transition metals, invoking a
simplified model of rectangular $d$-density of states due to Friedel\cite{Friedel}, 
the bond energy term is sometimes approximated simply via the second moment
of the density of states \cite{gupta,mukherjee,sutton,pettifor}.

 An approach that has been widely used in this context is the 
 embedded atom method (EAM) of Daw and Baskes\cite{Daw2,Daw3},
 where the energy of the metal
 is viewed as the energy to embed an atom into the local electron density provided 
 by the remaining atoms of the system, plus a sum of pair interaction potentials
 between the atoms. This method, which was developed almost simultaneously 
 by Finnis and Sinclair\cite{Finnis26}, starts with the ansatz:  
\begin{equation} 
E_{tot}=\sum_{i}F_{i}(\rho_{h,i})+\frac{1}{2}{\sum_{\substack{i,j\\(i\neq j)}}
\Phi (R_{ij})},
\label{etot} 
\end{equation} 
\begin{equation}
\rho_{h,i}=\sum_{j(j \neq i)} f_j(R_{ij}),
\end{equation}
where $E_{tot}$ is the total internal energy, $\rho_{h,i}$ is the host electron 
density at atom $i$ due to all other atoms, $f_j$ is the electron density of 
atom $j$ as a function of distance from its center, $R_{ij}$ is the separation 
distance between atoms $i$ and $j$. The function $F_i(\rho_{h,i})$, called the 
embedding energy, is the energy to embed atom $i$ in an electron density 
$ \rho_{h,i}$. $\Phi_{ij}$ is a two-body central pair potential between atoms 
$i$ and $j$. The host electron density $\rho_{h,i}$ is assumed to be a linear 
superposition of spherically symmetric contributions from all individual atoms 
except atom $i$. 

In the model proposed by Finnis and Sinclair\cite{Finnis26}  the total energy  
of a system of atoms is assumed to consist of a binding term proportional to the 
square-root of the local density and a repulsive pairwise potential term.   
Another approach to this  type of theory was provided by Manninen\cite{Manninen27}, 
Jacobsen, N$\phi$rskov, and Puska\cite{Jacobsen28}, who derived the functional 
form of the  two terms in Eq.(1) by using the density functional theory.  
Although Daw and Baskes\cite{Daw2,Daw3} had also used the density functional 
theory to justify the use of  Eq.(\ref{etot}), in most applications of the method 
the two terms of Eq.(\ref{etot}) were written in terms of parameters that were 
fitted to observed properties of the system.  In this sense, the EAM remains an 
empirical or at best a semi-empirical method. However, the simplicity and ease 
with which it could be applied to a large variety of situations has led to its 
wide use in the study of liquids\cite{Foiles6,Mei30,Mei-Al}, 
alloys\cite{Foiles8,Johnson9}, surfaces and interfaces
\cite{Daw5,Nelson17,Nelson18,Daw11,Daw11b,Felter13,Daw14,Daw15,Foiles16,Johnson4}, 
impurities and other defects in solids\cite{Daw2,Daw3,Bas10}. Relevant to the 
present work is the study by Foiles and Adams\cite{foiles-adams},
who investigated some thermodynamic properties of six fcc metals: 
Cu, Ag, Au, Ni, Pd and Pt via quasiharmonic calculations and molecular dynamics 
simulation using the EAM potentials developed by Foiles, Baskes and 
Daw\cite{Foiles8}.

By replacing the atomic electron density with an exponentially  decaying function, 
Johnson\cite{Johnson4} developed a set of analytic EAM functions for the nearest 
neighbor model of fcc metals. However, this model had the limitation that all 
materials were forced to have the same anisotropy ratio of the shear moduli
: $c_{44}/(c_{11}-c_{12})=1$.  Oh and Johnson\cite{OhJohnson} extended the 
model beyond nearest neighours, at the cost of sacrificing the simple analytic 
form.  Mei {\sl et al.} \cite{Mei1} have overcome this challenge and extended 
this nearest-neighbor model into the one in which the embedded atom potentials  
are analytic and valid for any choice of the cut-off distance. In particular 
they derived the values of the parameter for their EAM potential and density 
functions for six fcc metals: Ni, Pd, Pt, Cu, Ag, and Au.
 They used  this EAM potential in a molecular dynamics study of thermal
 expansion and specific heat of liquid Cu as a function of temperature, using 
 a cut-off for the potential and embedding functions somewhere between the 
 third and fourth neighbors for the corresponding crystalline case. In  later 
 studies they applied the model to study self-diffusion in the liquid phase of 
 the above six metals\cite{Mei30} and the melting in Al\cite{Mei-Al}. Kuiying 
 {\it et al.}\cite{kuiying} have used this EAM potential in a molecular dynamics 
 study of the local structure in  supercooled liquid and solid Cu and Al.
Although Mei {\sl et al.} \cite{Mei1} had derived the values of the EAM 
parameters for six fcc metals: Ni, Pd, Pt, Cu, Ag, and Au, to this date a 
systematic study of the vibrational and thermodynamic properties of these 
solids in the bulk crystalline phase using these parameters has not been 
carried out.  This work presents such a study  for the first time and explores 
to what extent this particular EAM model is successful in reproducing the 
experimentally observed phonon spectra and thermodynamic properties for the 
above six bulk fcc metals.

The organization of this paper is as follows: in section \ref{model} we 
discuss some details of the EAM potentials of Mei {\sl et al.} \cite{Mei1} 
and its implementation in the present work.  In section \ref{phonons} we 
discuss the calculated phonon spectra for the six fcc metals and their
agreement with the experimental results. In section \ref{thermodynamic}  
we discuss the calculated and   the corresponding experimental values of 
various properties: thermal expansion coefficients, isothermal and adiabatic 
bulk moduli, specific heats at constant volumes and pressures, the Debye 
temperature and the Gr\"uneisen parameter as a function of temperature. In 
section \ref{conclusions} we summarize our results and conclusions about the 
validity of this particular EAM model.

\section{The Model}
\label{model}
 
The embedding function $F(\rho)$ and the two-body potential $\Phi(R)$ in 
Eq. (\ref{etot}) of Mei {\sl et al.} are given by:
\begin{eqnarray}
F(\rho)&=&-E_c{\bigg [}1-\frac{\alpha}{\beta} {\rm ln\/}\bigg 
[{\frac{\rho}{\rho_e}}{\bigg ]\bigg ]\bigg [} {\rho\over\rho_e} 
{\bigg ]}^{ \alpha/ \beta} \nonumber \\ &+& {1\over2}\phi_e
\sum_\Lambda s_{\Lambda}{ \rm exp\/}[-(p_\Lambda-1)\gamma] \nonumber \\
& & \times {\bigg [} 1+(p_\Lambda-1)\delta - p_\Lambda {\frac{\delta}
{\beta}} {\rm ln}{\bigg [}{\rho\over\rho_e}{\bigg ]\bigg ]}  \nonumber \\ 
& &\times {\bigg [ {\rho\over\rho_e} \bigg  ]}^{ p_\Lambda { \gamma\over\beta } },
\label{ff}
\end{eqnarray}

\begin{equation}
\Phi(R)=-\phi_e[1+\delta (R/R_{1e}-1)] {\rm exp\/}[-\gamma(R/R_{1e}-1)],
\label{fp}
\end{equation}
with  
\begin{equation}
\rho=\sum_{\Lambda}s_{\Lambda}f(R_{\Lambda}),
\label{rh}
\end{equation}
\begin{equation}
f(R)={f_e}\sum_{\tau=0}^{\kappa} {c_\tau}{({R_{1e} /R})}^\tau.
\label{fff}
\end{equation}
In Eq. (\ref{ff}),  $\alpha$ is defined by
\begin{equation}
\alpha=(9B_e\Omega_e/E_c)^{1/2},
\end{equation}
$\Omega$ is the atomic volume, $B$ the bulk modulus, $E_c$ the cohesive energy. 
The subscript $e$ refers to the equilibrium value. $s_{\Lambda}$ in 
Eq. (\ref{rh}) is the number of atoms on the $\Lambda$-th neighbor shell with 
respect to a given reference atom.  $p_{\Lambda}$ refers to the $\Lambda$-th 
neighbor shell via 
\begin{equation}
R_\Lambda={p_\Lambda}{R_1},      \Lambda=1,2,...,
\end{equation}
where $R_1$ is the distance of the first neighbor shell with respect to a 
reference atom and $R_{\Lambda}$ the distance of the $\Lambda$th-neighbor shell. 
The constants $p_\Lambda$ depend on the  crystal structure type:  for the fcc 
structure $p_\Lambda=\sqrt{\Lambda}$.

Mei {\sl et al.}\cite{Mei1} used $\kappa=5$ in Eq.(\ref{fff}) to fit the atomic 
charge density $\rho$.  Only ratios of electronic densities appear in Eq.(\ref{ff}) 
and hence $\rho_e$ cancels out.  The constant $f_e$ was set equal to $\rho_e/12$. 
The constant $\beta$ was taken from  previous works by Johnson\cite{Johnson4,Johnson9}.
The remaining constants were obtained by fitting to the measured values of unrelaxed 
vacancy formation energy and the elastic constants, all of which were calculated by 
using three shells of neighbors for all the fcc metals considered. The exact details 
of the fitting procedure used is given in Ref. [\onlinecite{Mei1}].

The values of the constants thus generated and given in Table I of Mei 
{\sl et al.}\cite{Mei1}, were used in the present work with one notable difference.  
Mei {\sl et al.}\cite{Mei1} obtained the values of the parameters in their model 
by considering three shells of neighbors for the fcc solid. In the molecular 
dynamics study of liquid and solid Cu, they used a cut-off distance in their 
embedding function and pair-potential lying between the third and the fourth 
nearest neighbors. We have calculated the elastic constants $C_{11}$, $C_{12}$ 
and $C_{44}$ using the homogeneous deformation method and also from the long 
wavelength phonons and studied their variation with respect to the number of 
neighbor shells included in the calculation. For three shells of neighbors 
our results obtained via the homogeneous deformation method agrees well with 
those obtained by Mei {\sl et al.}\cite{Mei1}.  However, the values obtained 
via the homogeneous deformation method and from long wavelength phonons differ 
significantly from each other as well as from the experimental values. The  
differences in the results obtained by the two methods as well as between the 
calculated and experimental values, decrease on increasing the number of 
neighbor shells in the calculation and  practically disappear as the number 
of neighbor shells reaches six. Based on this result, we have calculated the 
phonon spectra and all the physical properties by using six shells of 
neighbors for this model of EAM. We have compared the phonon frequencies for 
the test case of Cu by using three to six shells of neighbors 
The maximum difference, which is found for low energy phonons, between the 
three and six neighbor shells is of the order of 1\%.  The differences in 
the phonon frequencies between the three and six neighbor shell calculations 
decrease with increasing wave vector, and at the zone boundary they are of 
the order of .01\%. The static energy of the solid, given by Eq.(\ref{etot}), 
changes by $\sim 0.7$\% as the number of shells increases from three to six. 
However, there is virtually no change in the location of the minimum in the 
static energy as a function of lattice parameter, as well as the curvature 
at the location of the minimum. 
\section{Phonons}
\label{phonons}

The phonon spectra were calculated as usual by diagonalizing the dynamical matrix, 
obtained from the  Fourier transform of the  force constant tensor 
${\bf \Phi}_{ij}(l,m)$ given by 
\begin{equation}
{\bf \Phi}_{ij}(l,m)={\partial^2 E_{tot}\over {\partial R^i(l)\partial R^j(m)}},
\end{equation}
where $l,m$ are the labels of the atoms. For $E_{tot}$ given by Eqs.(1) and (2), 
and with  $l\neq m$  
\begin{eqnarray}
{\bf \Phi}_{ij}(l,m)&=&-F_l'(\rho _l)f_m''(R_{lm}){R_{lm}^{j}R_{lm}^{i} 
\over R_{lm}^2} \nonumber 
\\ & &-F_m'(\rho _m)f_l''(R_{lm}){R_{lm}^{j}R_{lm}^{i} 
\over R_{lm}^2} \nonumber \\ & &-\Phi''(R_{lm})
{R_{lm}^{j}R_{lm}^{i} \over R_{lm}^2} -[F_l'(\rho_l)f_m'(R_{lm}) 
\nonumber \\ & &+F_m'
(\rho_m)f_l'(R_{lm})+\Phi '(R_{lm})] \nonumber \\ & &\times 
{\bigg \{}{\delta_{ij}\over R_{lm}} -{R_{lm}^{i}R_{lm}^{j}
\over R_{lm}^3}{\bigg \}}+\sum_{n(\neq  l,m)}F_n''(\rho_n) 
\nonumber \\ & &\times f_l'(R_{ln})f_m'(R_{mn}) {R_{mn}^{j}
\over R_{mn}}{R_{ln}^{i}\over R_{ln}},
\label{Fi} 
\end{eqnarray}
where the prime in the function in the above equation denotes the derivative 
of the function with respect to its argument.  Though written somewhat differently 
the above expression is in agreement with that derived by Finnis and 
Sinclair\cite{Finnis26}.

\begin{figure}
\renewcommand{\baselinestretch}{1}
\includegraphics[width=8.6cm]{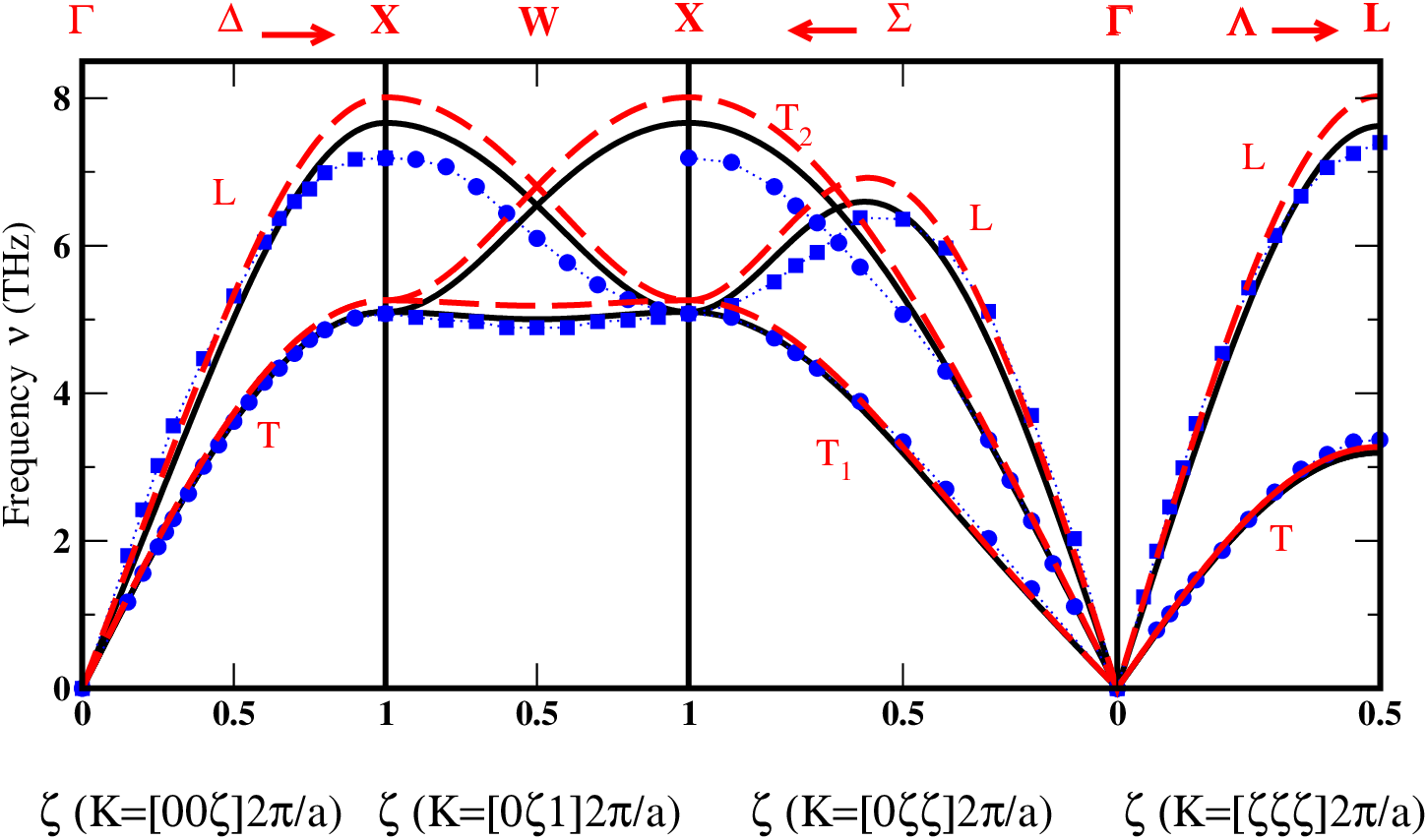}
\vspace{0.25in}
\caption{Phonon dispersion curves for Cu. The solid lines are the calculated 
phonon dispersion 
curves at the room temperature equilibrium lattice parameter of 
3.6131 \AA\cite{Kit53}.  The square and round points 
are the experimental data  at $296{\rm K}$ from Ref. [\onlinecite{EC34}]. 
$L$ and $T$ represent 
transverse and longitudinal modes, respectively.}
\label{Cugra}
\end{figure}

\begin{figure}
\renewcommand{\baselinestretch}{1}
\includegraphics[angle=270,width=8.6cm]{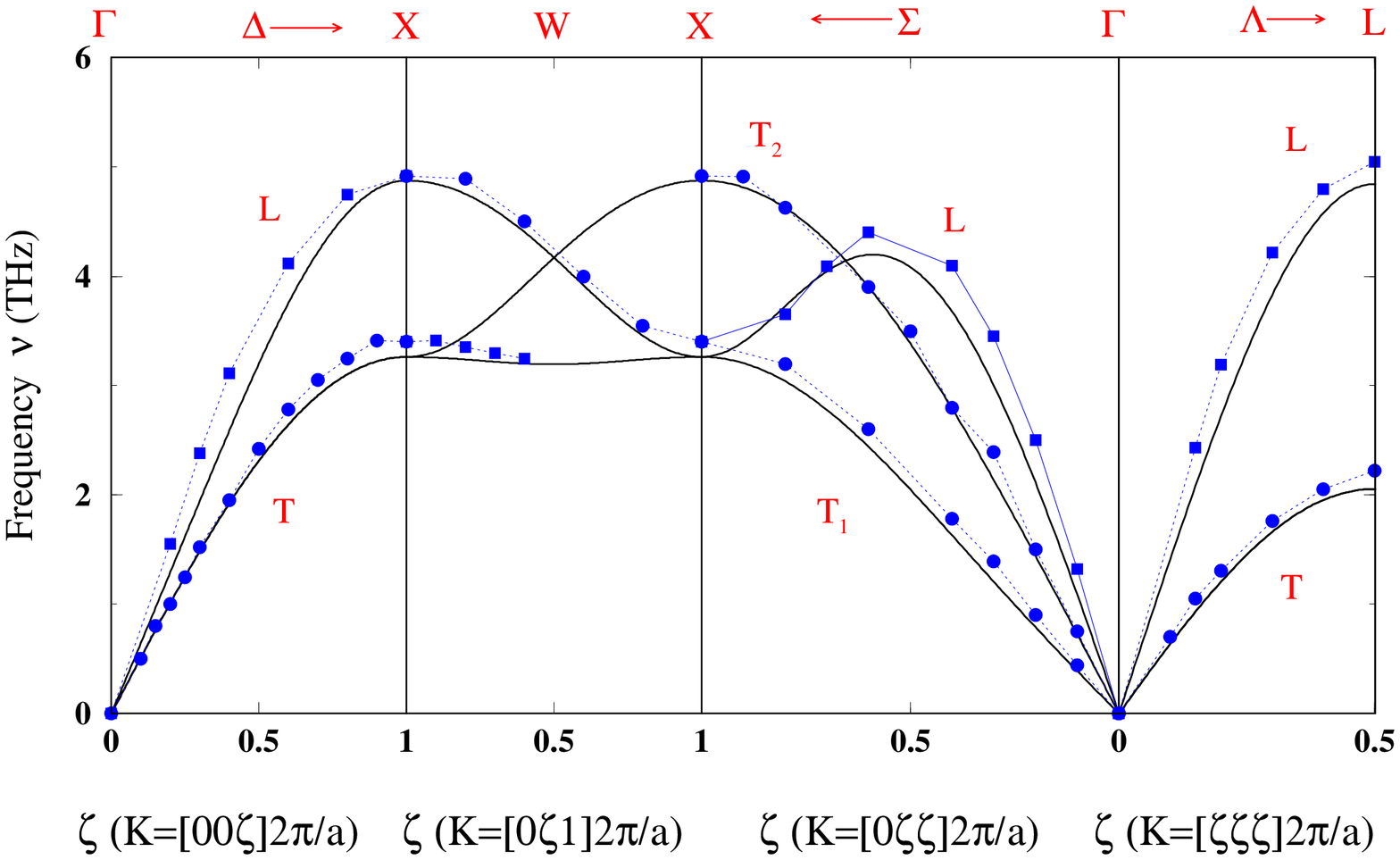}
\vspace{0.25in}
\caption{Phonon dispersion curves for Ag. The solid lines are the calculated phonon
dispersion curves at the room temperature equilibrium lattice parameter\cite{Kit53}. 
The square and round points are the experimental data from Ref.[\onlinecite{WA33}] 
at room temperature.  $L$ and $T$ represent
transverse modes and longitudinal modes respectively.}
\label{Aggra}
\end{figure}

\begin{figure}
\renewcommand{\baselinestretch}{1}
\includegraphics[angle=270,width=8.6cm]{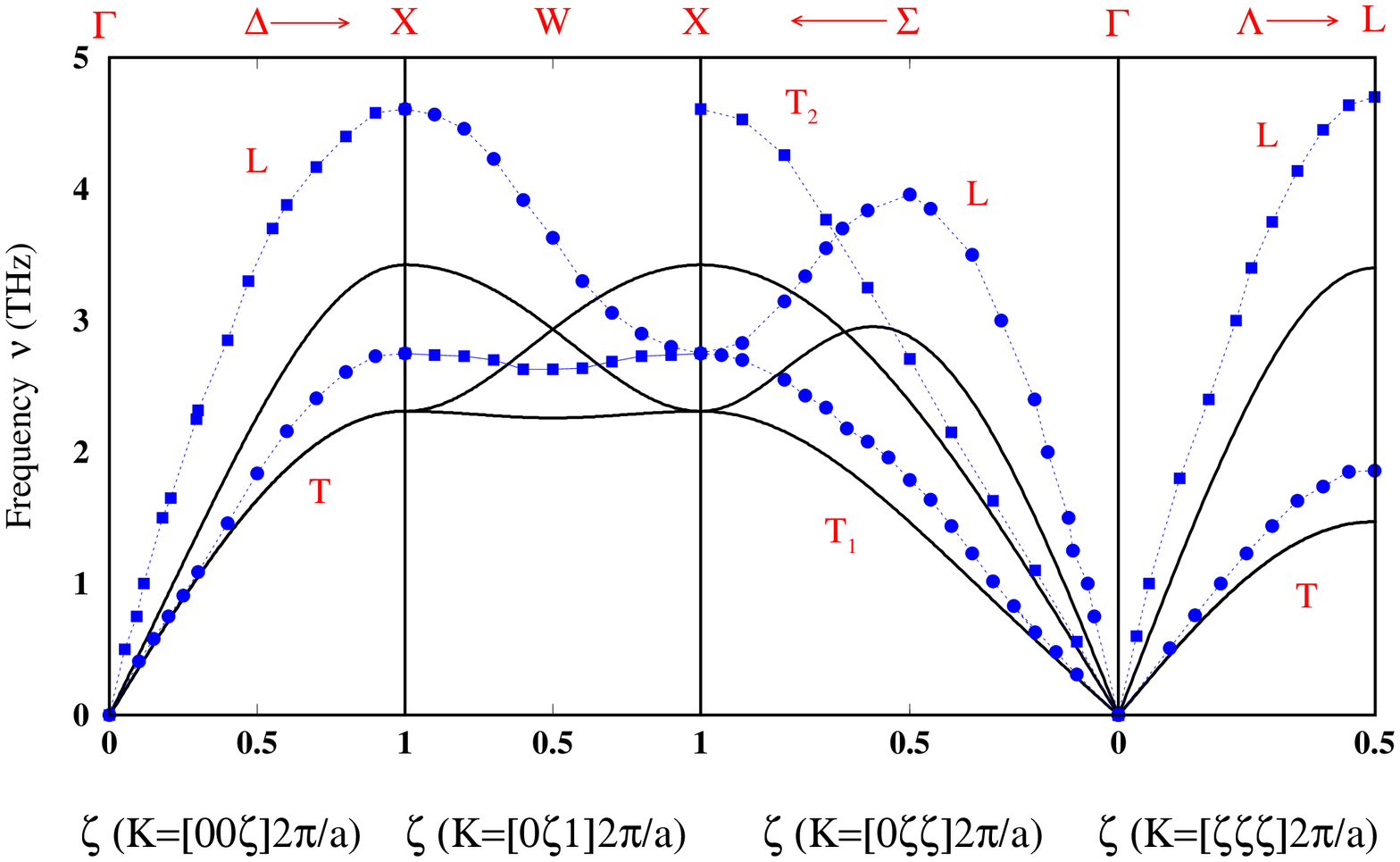}
\vspace{0.25in}
\caption{Phonon dispersion curves for Au. 
The solid lines, square and round points
and the symbols $L$ and $T$ have the same meanings as in Figs. 1 and 2. 
The experimental data at $296{\rm K}$ is  from Ref. [\onlinecite{Lynn38}]}
\label{Augra}
\end{figure}

\begin{figure}
\renewcommand{\baselinestretch}{1}
\includegraphics[angle=270,width=8.6cm]{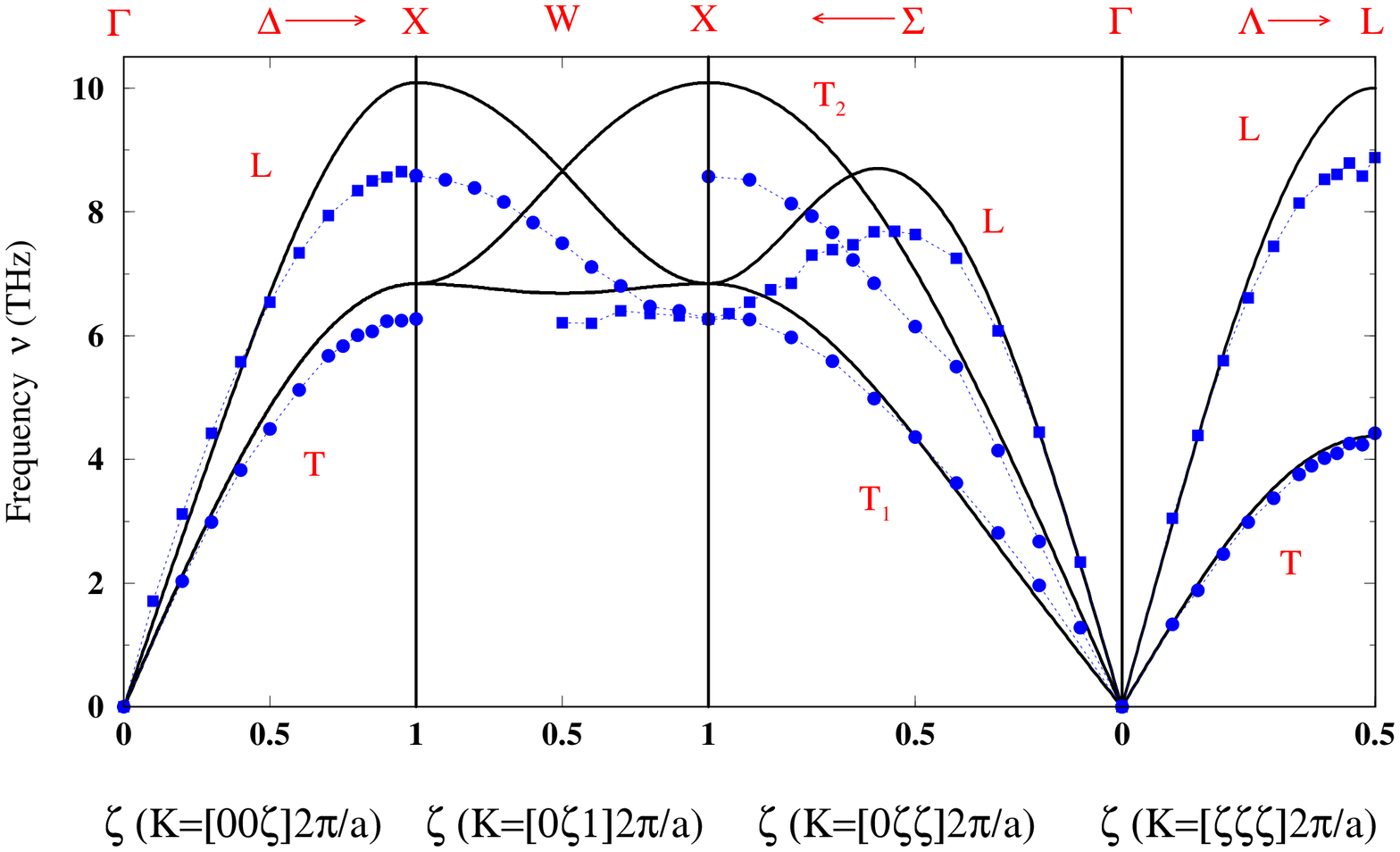}
\vspace{0.25in}
\caption{Phonon dispersion curves for Ni. The solid lines, square and round points
and the symbols $L$ and $T$ have the same meanings as in Figs. 1 and 2.  
The experimental data taken at $296{\rm K}$ is from Ref.[\onlinecite{Birgr37}].}
\label{Nigra}
\end{figure}

\begin{figure}
\renewcommand{\baselinestretch}{1}
\includegraphics[angle=270,width=8.6cm]{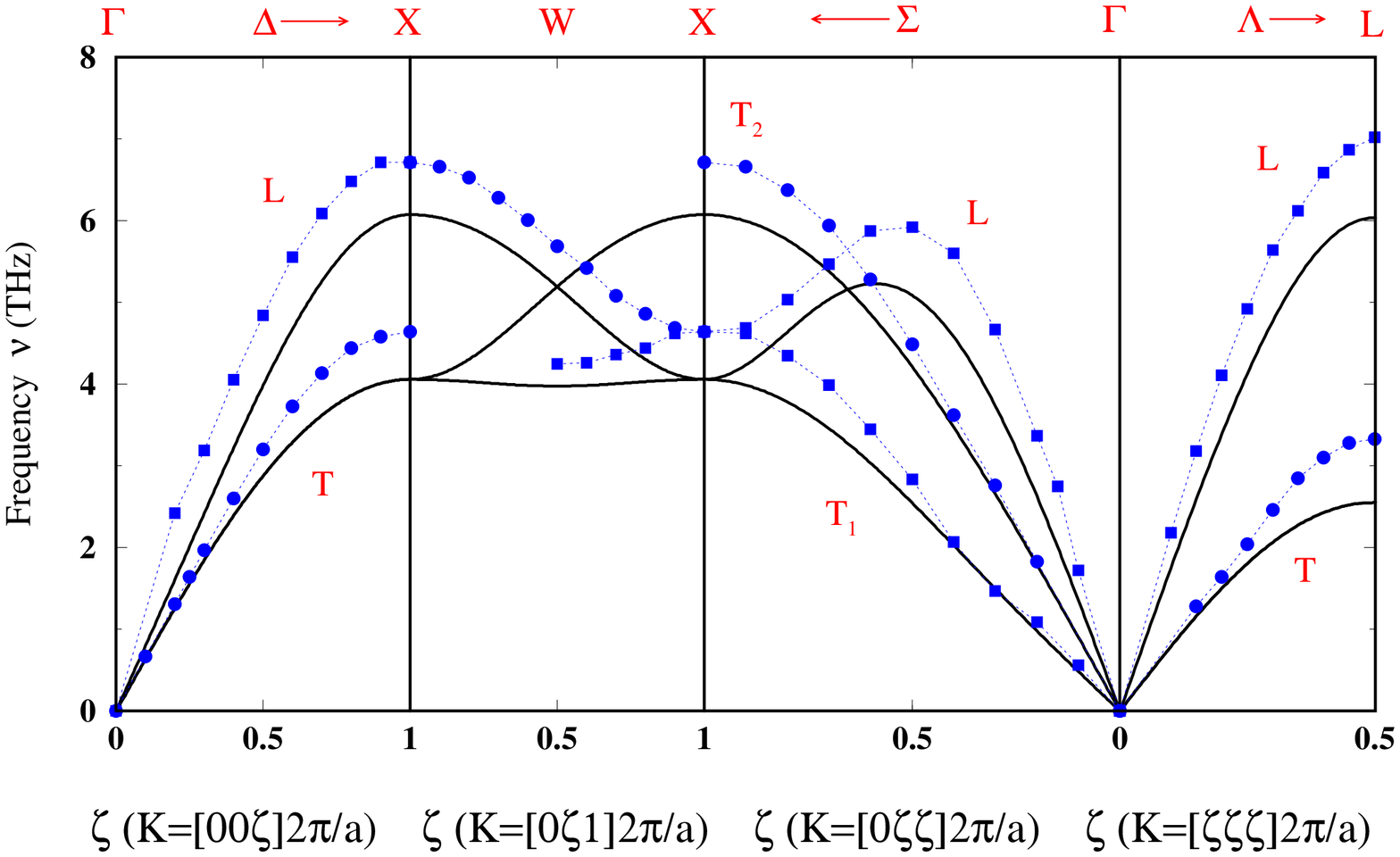}
\vspace{0.25in}
\caption{Phonon dispersion curves for Pd. 
The solid lines, square and round points
and the symbols $L$ and $T$ have the same meanings as in Figs. 1 and 2.  
The experimental data taken at $120{\rm K}$ is from Ref. [\onlinecite{Miller35}].}
\label{Pdgra}
\end{figure}

\begin{figure}
\renewcommand{\baselinestretch}{1}
\includegraphics[angle=270,width=8.6cm]{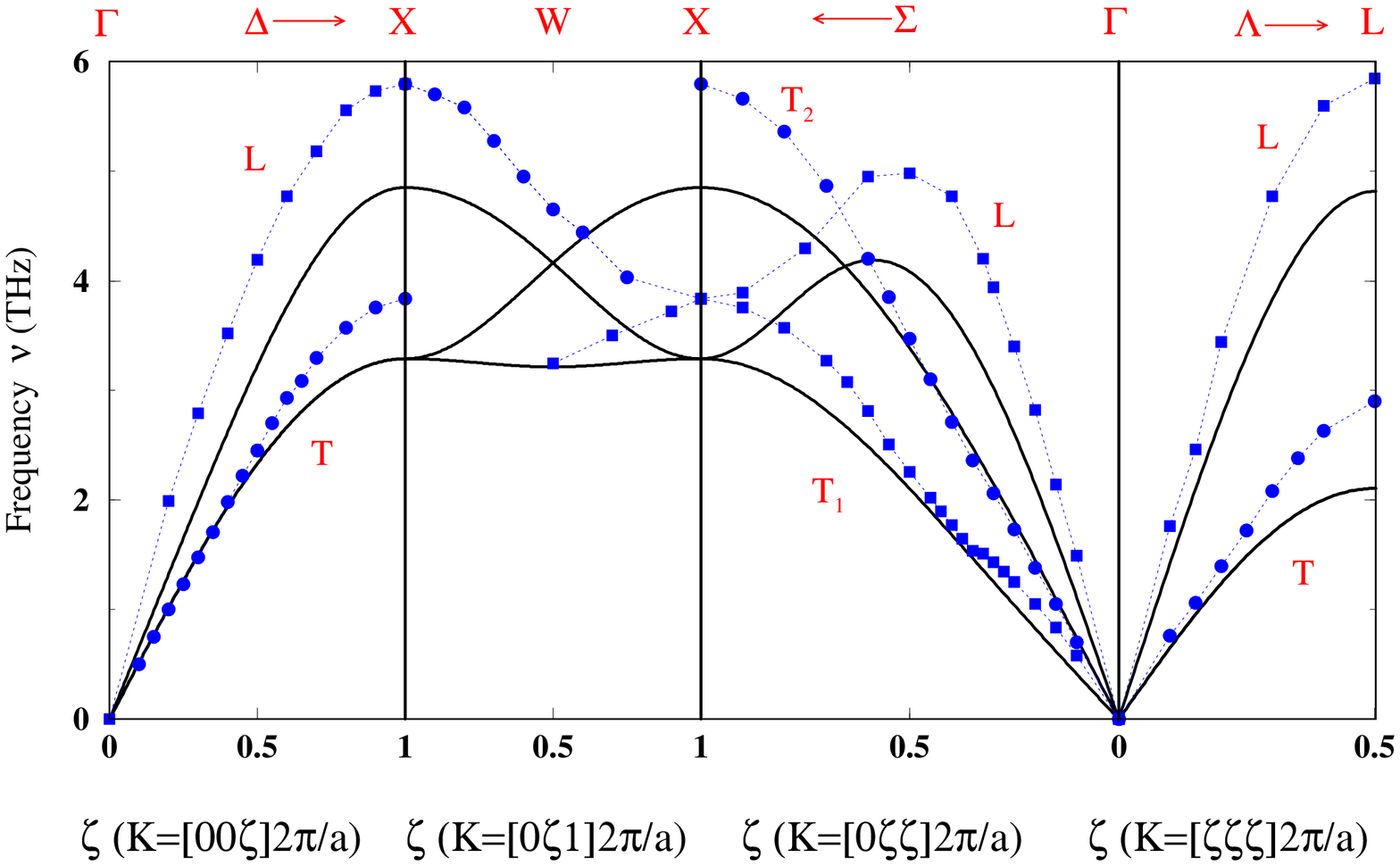}
\vspace{0.25in}
\caption{Phonon dispersion curves for Pt. The solid lines, square and round points
and the symbols $L$ and $T$ have the same meanings as in Figs. 1 and 2.
The experimental data is from Ref.[\onlinecite{Dutton36}].}
\label{Ptgra}
\end{figure}

The  calculated phonon frequencies for the six fcc metals Ni, Pd, Pt and Cu, Ag, 
Au are plotted in Figs.(\ref{Cugra}-\ref{Ptgra}).  
In Fig. (\ref{Cugra}) we compare the calculated phonon spectrum in Cu with the 
experimental results of Svensson, Brockhouse and Rowe\cite{EC34}. The squares 
and the circles represent the phonon frequencies from the inelastic neutron 
scattering experiment at room temperature (296 K), while the solid line 
denotes the calculated spectrum. For estimating the importance of the embedding 
term we have used dashed lines to show the frequencies obtained by considering 
only the pair potential term, i.e., neglecting the contribution from 
the embedding function. The pair potential and the embedding terms contribute 
to the force constant with opposite signs: without the embedding term the phonon 
frequencies at high values of the wave vectors would have much worse agreement 
with the measured frequencies.  We find that the three-body terms coming from 
the embedding function (the last term in Eq.(\ref{Fi})) have a negligible effect 
on the phonon frequencies, 
the values obtained with and without these terms are virtually indistinguishable. 
We have also compared the calculated phonon frequencies with the 
experimental values for off-symmetry wave vectors, as given by Nilsson and 
Ronaldson\cite{nielsson}. The agreement is similar to that for the wave vectors 
along the symmetry directions, in the sense that the agreement is very good for 
small wave vectors, while becoming progressively worse with increasing wave 
vectors to the same extent as for the symmetry direction wave vectors shown in 
Fig. (\ref{Cugra}).

The term involving the embedding function in Eq.(\ref{etot}) contributes about 
$\frac{1}{4}$ to $\frac{1}{3}$ of the total energy for the model of Mei 
{\sl et al.} \cite{Mei1}, but their contribution to vibrational frequencies is 
less than 10\%. Hence, the embedding term contributes to the temperature variation 
of thermodynamic properties mainly via the static part of the free energy.

For Cu the phonon frequencies calculated by considering only the contributions 
from the nearest neighbors in the fcc structure produce about 90\% of the final 
converged frequencies originating from all neighbors. This is in agreement with 
the observation of Svensson {\it et al.} \cite{EC34}, who used several 
Born-von K\'{a}rm\'{a}n force constant models involving neighbors up to various 
different shells to study the agreement between the calculated phonon frequencies 
and those from the inelastic neutron scattering experiment. Their least square 
fits of atomic  and planar force constants to the observed phonon frequencies 
indicate that the nearest neighbor forces dominate in Cu, although longer range 
forces extending at least to the sixth-nearest neighbors are needed for a 
complete agreement between the calculated and experimental frequencies. 
A similar result was found by Nilsson and Ronaldson\cite{nielsson} in a neutron 
crystal spectrometer study of phonon frequencies in Cu at wave vectors of both 
symmetry and off-symmetry points in the Brillouin zone at a temperature of 80 K.

Among the six fcc metals, the agreement between the calculated and experimental 
phonon spectra is almost perfect for Ag and only slightly less so for Cu. For 
all other metals the agreement becomes increasingly worse with increasing wave 
vectors in all the symmetry directions, the worst case being that of Au. For Pd, 
Pt and Au the calculated phonon frequencies are underestimated with respect to 
the experimental results, while they are somewhat overestimated for Ni.
For Cu\cite{Nelson17}, Ag\cite{Nelson18}, Ni and Pd\cite{Daw5}, this trend as 
well as the level of agreement between the calculated and experimental phonon 
frequencies is the same as in the EAM scheme of  Daw and Baskes\cite{Daw3}, 
where the total charge density at an atomic location is calculated from 
the {\sl ab initio} Hartree-Fock results for free atom charge densities 
\cite{clementi,Mclean}. As pointed out by Daw and Hatcher\cite{Daw5}, 
as long as the fitting is done primarily to elastic constants, which involve 
phonons near the zone center  only, the agreement for high frequency phonons 
near the zone boundary is not guaranteed. In this respect, phonons calculated 
via {\it ab initio} electronic structure methods, which can capture the details 
of Fermi surface topology without resorting to any empirical 
fitting procedure, can yield superior
results\cite{savrasov,papa2}, although results do vary depending on the method 
of electronic structure calculation and details of the implementation of 
exchange-correlation potentials, etc\cite{sobhana}.
Of course if one is simply interested in the phonon spectra, Born-von 
K\'{a}rm\'{a}n fit to force constants can yield
very satisfactory results\cite{EC34}. Note that the agreement with the 
experimental phonon spectra for Cu obtained by
Cowley and Shukla\cite{shukla} by using a nearest neighbor Born-Mayer 
potential with volume-dependent prefactor is 
as good as obtained in the present EAM model. The total energy of the crystal 
in this study by Cowley and Shukla\cite{shukla}
consisted of kinetic, exchange and correlation energies of the electron gas 
and an electron-ion interaction term
in addition to the nearest neighbor Born-Mayer potential.  

\section{Thermodynamic properties}
\label{thermodynamic}

In the quasiharmonic approximation, the total Helmholtz free energy of the crystal 
at temperature $T$ and volume $V$ or lattice constant $a$ is given by
\begin{eqnarray}
F(a,T)&=&E_{tot}(a)+F_{vib}(a, T) \nonumber \\ &=&E_{tot}(a) + k_{B}\sum_{{\bf K}s} 
\nonumber 
\\ & & \times {\rm ln}\{2sinh{\hbar \omega_{s}({\bf K},a)\over 2k_{B}T}\big \}
\label{Fa}
\end{eqnarray}
where $E_{tot}$ is the static total energy given by  Eq.(\ref {etot}) at a given 
volume $V$ or lattice constant $a$. $\omega_{s}({\bf K},a)$ is the frequency of 
$s$th mode for a given wave vector ${\bf K}$ and lattice constant $a$.  
$\hbar$ and $K_{B}$ are the Planck's and Boltzmann's constants, respectively.
 
All thermodynamic properties are calculated from the free energy given by Eq.(\ref{Fa}).
In obtaining the wave vector  sum in Eq.(\ref{Fa}) we consider a $20^3$ uniform grid 
in the Brillouin zone (BZ) giving 256 wave vectors in the irreducible part. The sum 
is performed using these 256 wave vectors, appropriately weighted according to the 
point-group symmetry of the fcc solid. Temperature variation of thermodynamic 
properties is studied from $0K$  to $1400K$ in steps of $2K$.
  
At a given temperature $T$, the equilibrium lattice parameter is determined by the 
 minmum of the Helmholtz free energy, i.e.,
\begin{equation}
{\big (}{\partial F(a,T) \over \partial a}{\big )}_{T}={\partial E_{tot}(a) \over 
\partial a}+ {\big (}
{\partial F_{vib}(a,T) \over \partial a}{\big )}_{T}=0.
\label{pp}
\end{equation}

\begin{figure}
\vspace{0.25in}
\renewcommand{\baselinestretch}{1}
\includegraphics[angle=0,width=7.6cm, clip]{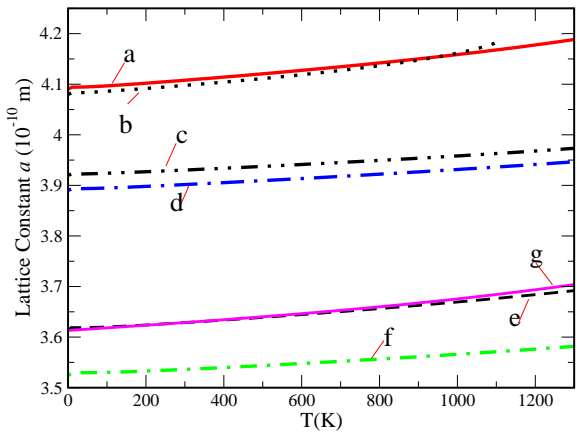}
\vspace{-0.05in}
\caption{Lattice Constant Against Temperature for the fcc Metals. Line a is for Ag, 
b is for Au, c is for Pt, d is for Pd, e is for Cu, and f is for Ni. Line g denotes 
the results of the MD simulation 
 Mei {\it et al.} [\onlinecite{Mei1}] for Cu.}
\label{LLL}
\end{figure}

The variation of lattice constant with temperature for the six fcc metals is shown 
in Fig.(\ref{LLL}).  It should be noted that our quasiharmonic results for Cu 
agree very well with those obtained by Mei {\sl et al.} \cite{Mei1} using molecular 
dynamics simulation. A small difference, increasing with
temperaure, appears above 900 K due to the anaharmonic effects not included in the 
quasiharmonic approach.  The results for all the six metals are as good as those 
obtained by Foiles and Admas\cite{foiles-adams}
using the EAM potentials of Foiles, Baskes and Daw\cite{Foiles8}

\begin{figure}
\renewcommand{\baselinestretch}{1}
\includegraphics[angle=0,width=8.0cm,clip]{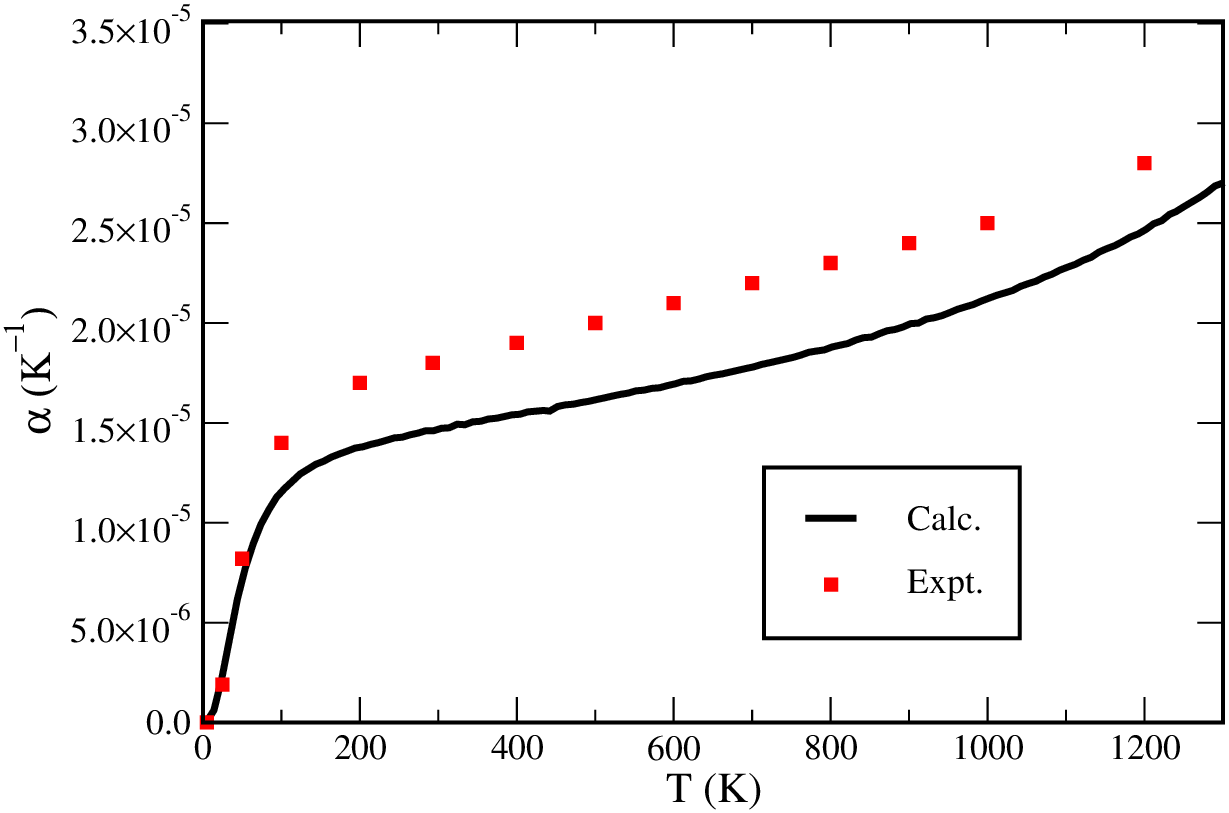}
\vspace{-0.05in}
\caption{Coefficient of linear thermal expansion $\alpha (T)$ as a function of 
temperature for Ag. The solid line denotes the calculated values, and the square 
points represent the experimental values from Ref. [\onlinecite{Tou43}].}
\label{alAg}
\end{figure}

\begin{figure}
\renewcommand{\baselinestretch}{1}
\includegraphics[angle=0,width=8.0cm,clip]{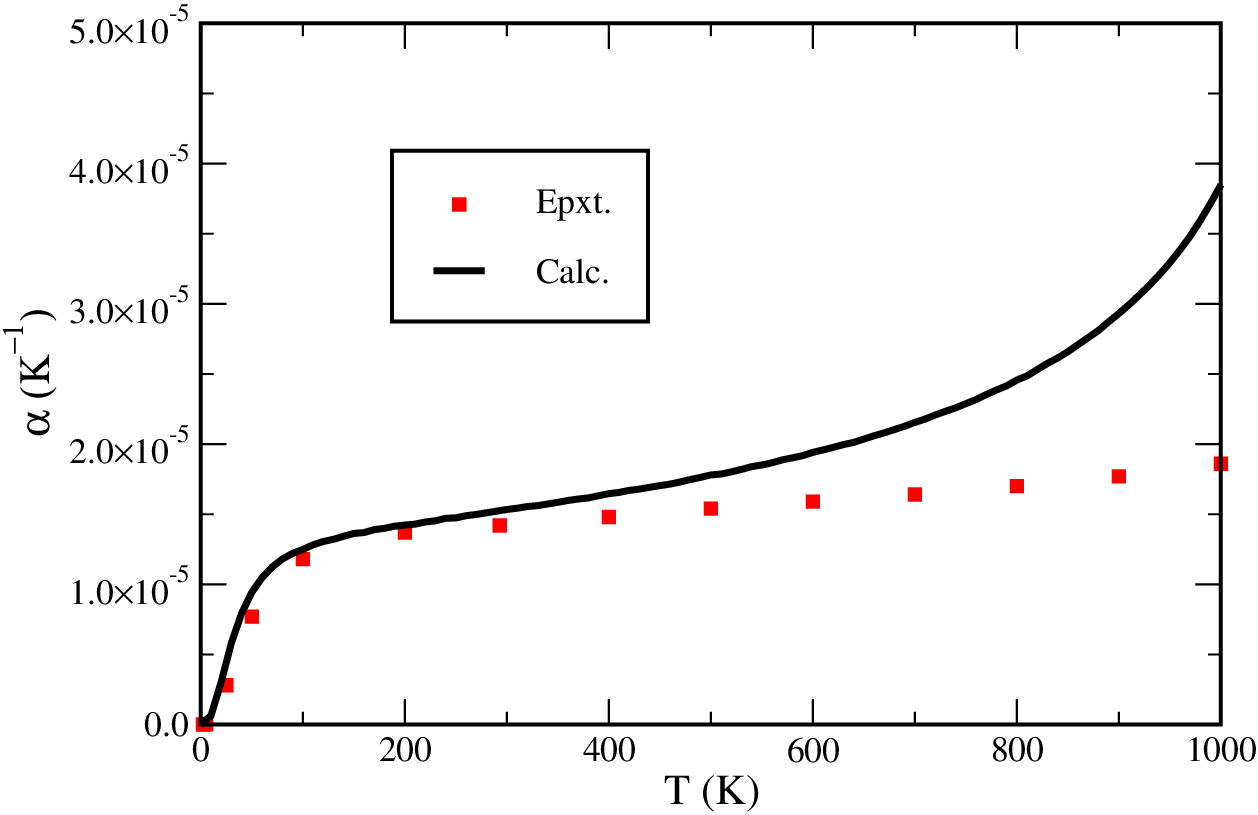}
\vspace{-0.05in}
\caption{Coefficient of linear thermal expansion $\alpha (T)$ as a function of 
temperature for Au. The solid line denotes the calculated values, and the square 
points represent the experimental values from Ref. [\onlinecite{Tou43}].}
\label{alAu}
\end{figure}

The coefficient of linear thermal expansion  is given by
\begin{equation}
\alpha (T)={1\over a_{e}(T)} {\big (}{d a_{e}(T)\over d T}{\big )}_p.
\label{al}
\end{equation}
In experimental works Eq.(\ref {al}) is often replaced with\cite{Mac44,shukla-mac}
\begin{equation}
\alpha (T)={1\over a_{e}(T_c)} {\big (}{d a_{e}(T)\over d T}{\big )}_p,
\label{all}
\end{equation}
where $T_{c}$ is a reference temperature, usually taken to be the room temperature.
To be consistent with  experimental and other theoretical work\cite{Mac44}, 
we take $T_{c}=293$.  Using Eq.
(\ref {all}), we calculate the coefficients of linear expansion $\alpha (T)$ for the 
six fcc metals and plot them in Figs.(\ref {alAg} - \ref {alAu}) along with the 
experimental values.  Thermal expansion coefficients for Cu, Ag and Ni calculated 
by Shukla and MacDonald\cite{shukla-mac} and MacDonald and MacDonald\cite{Mac44} 
based on  empirical nearest neighbor central force model and incorporating two 
lowest order, cubic and quartic, anharmonic effects show much better agreement with
experiment than those given by the present EAM model in the quasiharmonic 
approximation.
\begin{figure}
\renewcommand{\baselinestretch}{1}
\includegraphics[angle=0,width=8.0cm,clip]{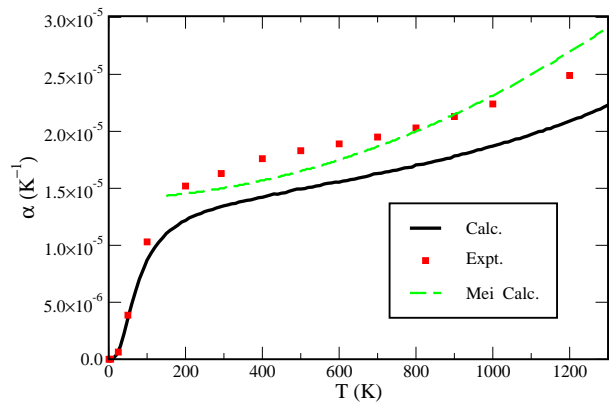}
\vspace{-0.05in}
\caption{Coefficient of linear thermal expansion $\alpha (T)$ as a function of 
temperature for Cu. the solid line is the calculated values, the square points 
are the experimental values from Ref. [\onlinecite{Tou43}], and the dashed 
line is the results of the MD simulation of Mei {\it et al.} [\onlinecite{Mei1}] }
\label{alCu}
\end{figure}

\begin{figure}
\renewcommand{\baselinestretch}{1}
\includegraphics[angle=0,width=8.0cm,clip]{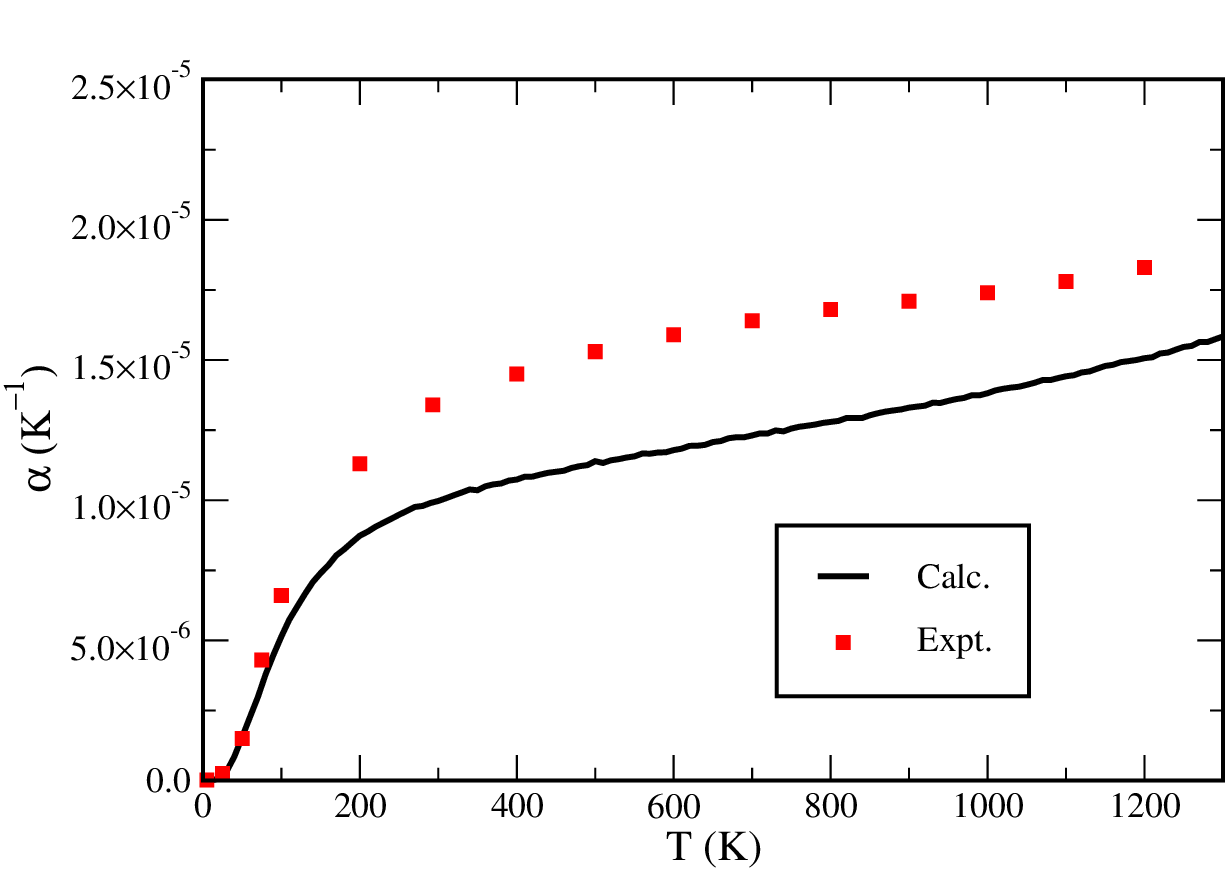}
\vspace{-0.05in}
\caption{Coefficient of linear thermal expansion $\alpha (T)$ as a function of 
temperature for Ni.  The solid line denotes the calculated values, and the square 
points represent the experimental 
values from Ref. [\onlinecite{Tou43}].}
\label{alNi}
\end{figure}
\begin{figure}
\renewcommand{\baselinestretch}{1}
\includegraphics[angle=0,width=8.0cm,clip]{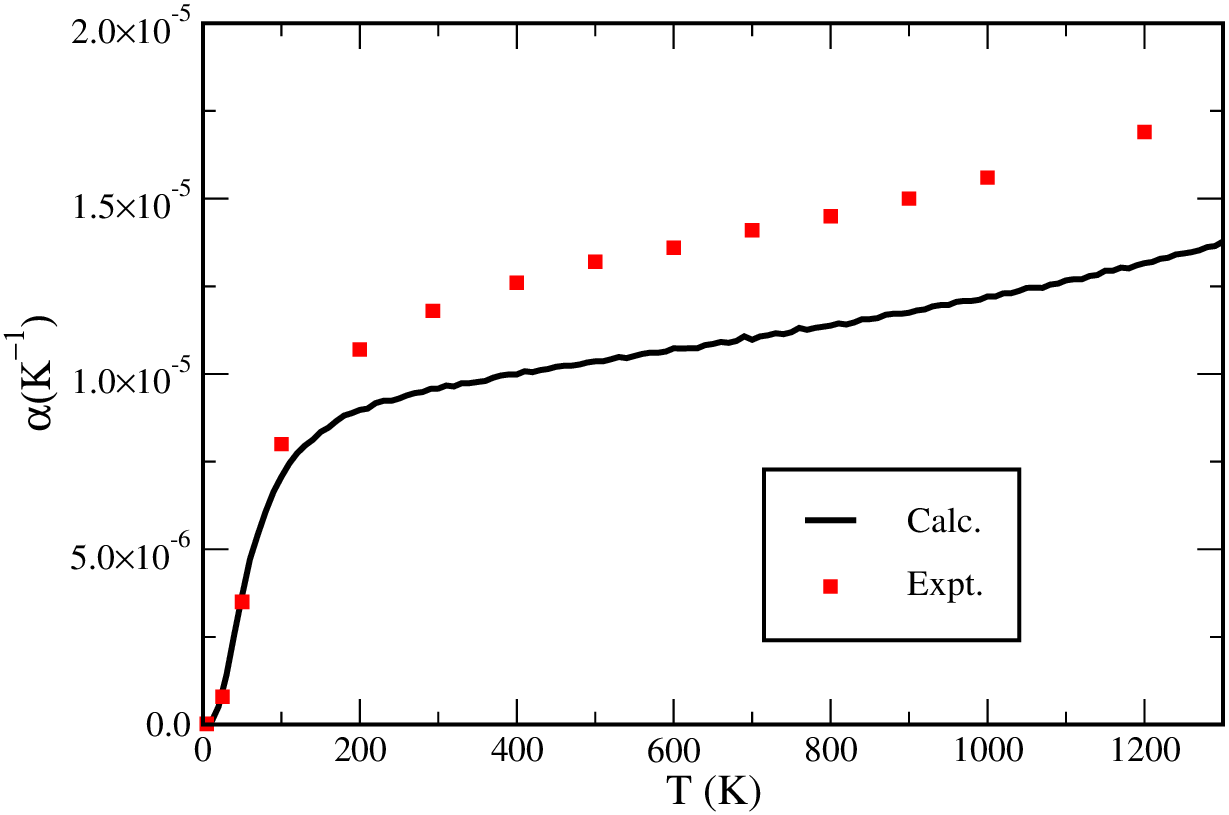}
\vspace{-0.05in}
\caption{Coefficient of linear thermal expansion $\alpha (T)$ as a function of 
temperature for Pd. The solid line denotes the calculated values, and the square 
points represent the experimental 
values from Ref. [\onlinecite{Tou43}].}
\label{alPd}
\end{figure}
\begin{figure}
\renewcommand{\baselinestretch}{1}
\includegraphics[angle=0,width=8.0cm,clip]{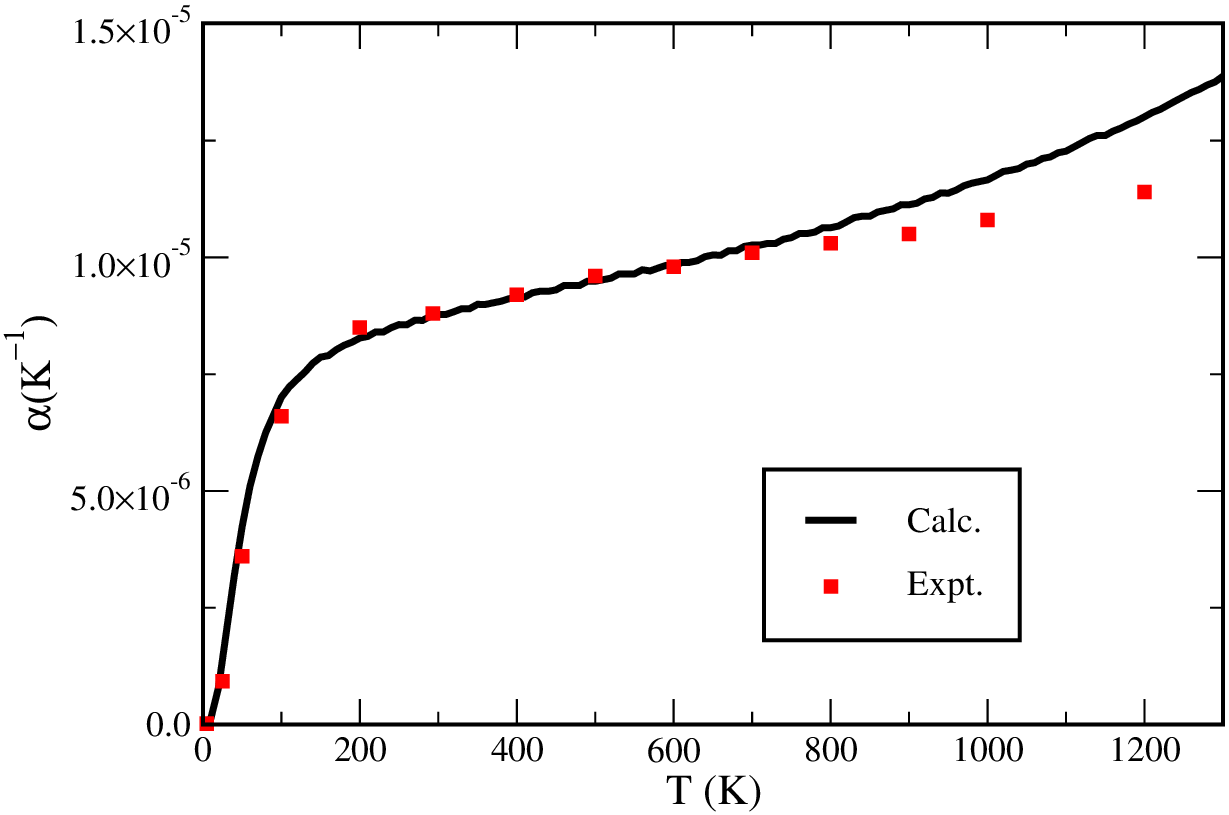}
\vspace{-0.05in}
\caption{Coefficient of linear thermal expansion $\alpha (T)$ as a function of 
temperature for Pt. The solid line denotes the calculated values, and the square 
points represent the experimental 
values from Ref. [\onlinecite{Tou43}].}
\label{alPt}
\end{figure}

The isothermal bulk modulus at a temperature T is  given by
\begin{equation}
B(T)=-V{\big (}{\partial P\over \partial V}{\big )}_{T}\;,\; 
P=-\frac{\partial F}{\partial V}.
\end{equation}
or 
\begin{eqnarray}
B(T)&=& V{\big (}{\partial^{2} F\over \partial V^{2}}{\big)}_{T}  \nonumber  \\
&=&V{\big (}{\partial^{2} E_{tot}\over \partial V^{2}}{\big)} 
+ V{\big (}{\partial^{2} 
F_{vib}(a,T)\over \partial V^{2}}{\big )}_{T}.
\label{B}
\end{eqnarray}

\begin{figure}
\renewcommand{\baselinestretch}{1}
\includegraphics[angle=0,width=7.6cm,clip]{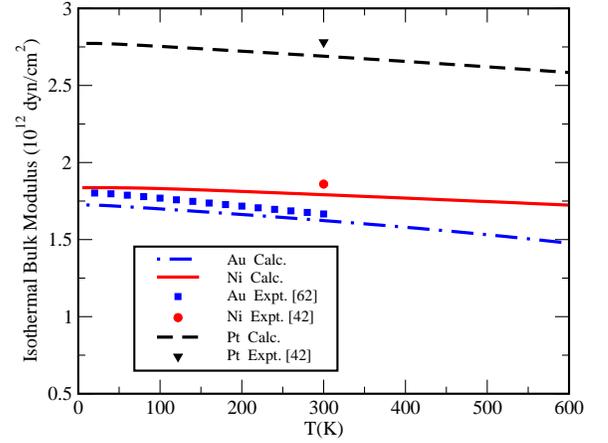}
\vspace{-0.05in}
\caption{Isothermal bulk modulus $B(T)$ as a function of temperature for Au, Ni 
and Pt.}
\label{BAu}
\end{figure}

\begin{figure}
\renewcommand{\baselinestretch}{1}
\includegraphics[angle=0,width=7.6cm,clip]{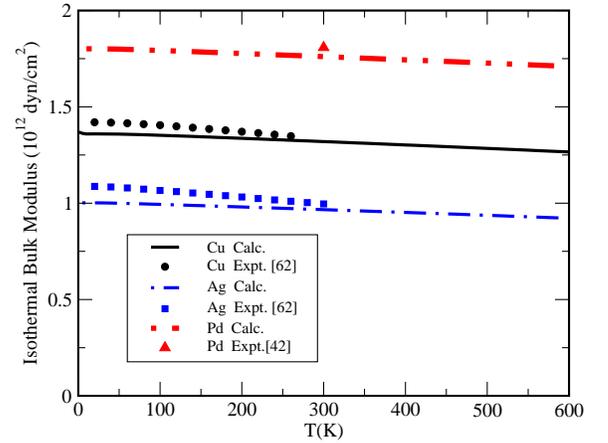}
\vspace{-0.05in}
\caption{Isothermal bulk modulus $B(T)$ as a function of temperature for Ag, 
Cu and Pd.}
\label{BCu}
\end{figure}

The second derivatives in  Eq.(\ref {B}) are obtained via numerical differentiation 
and  the results are compared with the available experimental data in Figs.
(\ref{BAu} - \ref{BCu}). Both the values and the trend, a small decrease with 
increasing temperature, are reproduced reasonably well in the calculation.

The specific heat at constant volume $C_{V}$ can be obtained from the temperature 
derivative of the total energy. However, the contribution  from the electrons 
excited across the Fermi level cannot be obtained from the EAM expression. Hence 
we write the specific heat at constant volume as
\begin{equation}  
C_{V}(T)=C^{ph}_{V}(T) + C^{el}_{V}(T). 
\end{equation}
where
the phonon part is obtained from the EAM model via
\begin{eqnarray}
C_{V}^{ph}(T)&=&\sum_{{\bf K},s} C_{V}({\bf K}s) \nonumber \\
&=&k_{B}\sum_{{\bf K}s}{\bigg \{}{\hbar \omega_{s}({\bf K})\over 2k_{B}T}{\bigg \}}^{2} 
\nonumber \\ & &\times {1\over sinh^{2}[\hbar \omega_{s}({\bf K})/2k_{B}T]}.
\label{cv}
\end{eqnarray}
We estimate the electronic part $C^{el}_{V}(T)$ via the well-known
expression:
\begin{equation}
 C^{el}_{V}(T) = \frac{\pi^2}{3}k_B^2N(E_F)T\;
\label{el-cv}
\end{equation}
where $N(E_F)$ is the electronic density of states at the Fermi level $E_F$. 
In the above expressions we consider the specific heats and $N(E_F)$  to be 
per atom. We use the well-known and well-documented Stuttgart TB-LMTO 
(Tight Binding Linear Muffin-tin Orbitals) code\cite{ole} to compute the values
of $N(E_F)$ for various lattice parameters.  Our results agree well with those 
given by other methods \cite{papa-MJW}. With increasing temperature, lattice 
parameter increases and electronic bands become narrower, resulting in a slight 
increase in $N(E_F)$. However, this increase is negligible and 
the variation of $C^{el}_{V}(T)$ with temperature remains essentially linear. 
We have actually calculated the $N(E_F)$ for 5-6 values of lattice parameters 
in the entire temperature range and used linear interpolation to find the values 
at all other lattice parameters.  Eq.(\ref{el-cv}), which is based on the 
independent fermionic quasiparticle picture\cite{grimvall,prange},
includes all the electron-ion, exchange and correlation effects as incorporated 
within the framework of density functional theory. What is neglected  
 is the electron-phonon interaction\cite{grimvall,prange}, which is presumably 
 small for most of the metals considered, as none of these exhibits 
 superconductivity down to almost absolute zero of temperature. 
 Theoretical calculations\cite{savrasov} of electron-phonon interaction show that
 for Cu it is indeed negligible, while for Pd it may not be so. Since the values 
 of the  electron-phonon coupling constant are known only for a few of these 
 metals, we have simply ignored this contribution.
 The experimental values contain the effects of electron-phonon interaction in 
 addition to the contributions from vacancies\cite{Mac44} and other defects.
 Note that in the high temperature limit the calculated quasiharmonic value of 
 $C^{ph}_{V}(T)$ reaches the classical harmonic value 3K$_B$ per atom, as it should.

We calculate the specific heat at constant pressure $C_{P}$ by using the relation: 
\begin{equation} 
C_{P}(T) - C_{V}(T)=-T{\big (}{\partial V\over \partial T}{\big )}^{2}_{P}{\big (}
{\partial P\over \partial V}{\big )}_{T}.
\label{Cp} 
\end{equation}
or 
\begin{eqnarray}
C_{P}(T) &= & C_{V}^{ph}(T) + C_{V}^{el}(T) \nonumber \\ & &+ 
{9\over 4}\alpha ^{2}(T) B(T)a_{e}^{3}(T)T.
\label{cvcpp}
\end{eqnarray}

\begin{figure}
\renewcommand{\baselinestretch}{1}
\includegraphics[angle=0,width=7.6cm,clip]{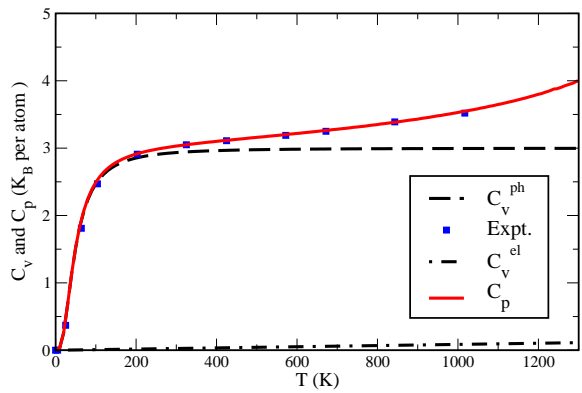}
\vspace{-0.05in}
\caption{Calculated temperature-dependence of heat capacity of Ag at constant volume 
$C_{V}^{ph}$, at constant pressure $C_{P}$ and the  electronic contribution 
$C_{V}^{el}$. 
The square points represent the experimental data for heat capacity at constant 
pressure from Ref.  [\onlinecite{Tou47}].}
\label{cpAg}
\end{figure}

\begin{figure}
\renewcommand{\baselinestretch}{1}
\includegraphics[angle=0,width=7.8cm,clip]{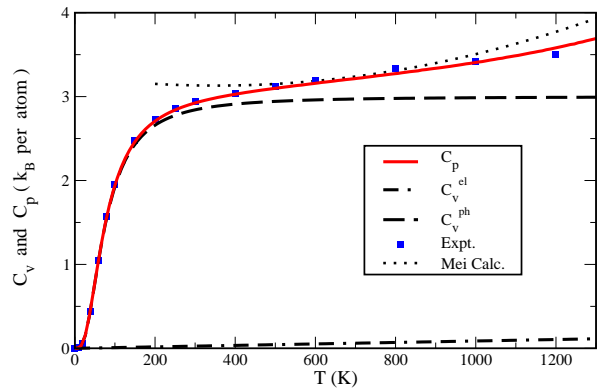}
\vspace{-0.05in}
\caption{Calculated temperature-dependence of heat capacity of Cu at constant volume 
$C_{V}^{ph}$, at constant pressure $C_{P}$ and the electronic contribution $C_{V}^{el}$.
 The square points are the experimental data for heat capacity at constant 
 pressure from Ref.  [\onlinecite{Tou47}].
 The dotted line denotes the results from the molecular dynamics simulation of 
 Ref. [\onlinecite{Mei1}].}
\label{cpCu}
\end{figure}

\begin{figure}
\renewcommand{\baselinestretch}{1}
\includegraphics[angle=0,width=7.6cm,clip]{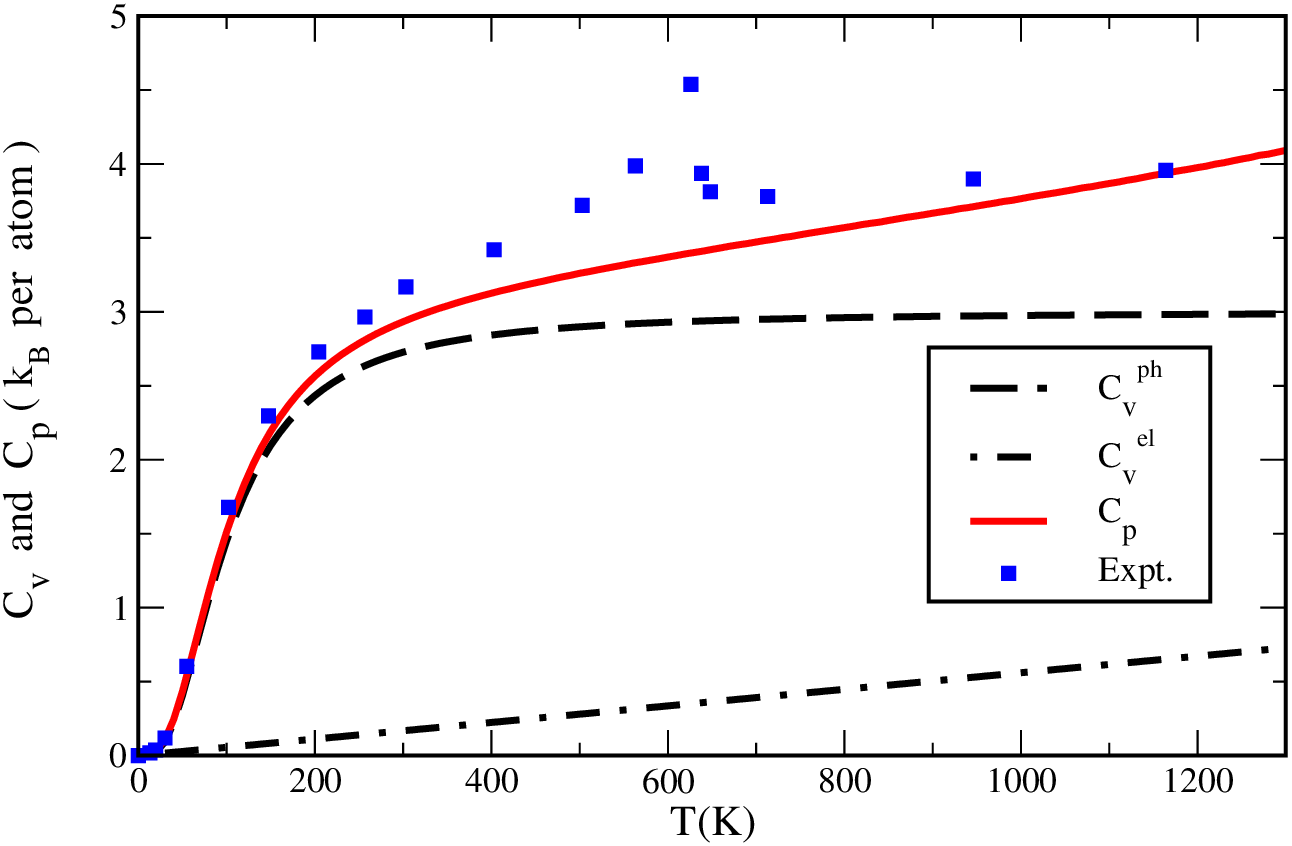}
\vspace{-0.05in}
\caption{Calculated temperature-dependence of heat capacity of Ni at constant volume
 $C_{V}^{ph}$, at constant pressure $C_{P}$ and the electronic contribution $C_{V}^{el}$. 
 The square points are the experimental data for heat capacity at constant pressure 
 from Ref.  [\onlinecite{Tou47}].}
\label{cpNi}
\end{figure}

\begin{figure}
\renewcommand{\baselinestretch}{1}
\includegraphics[angle=0,width=7.6cm,clip]{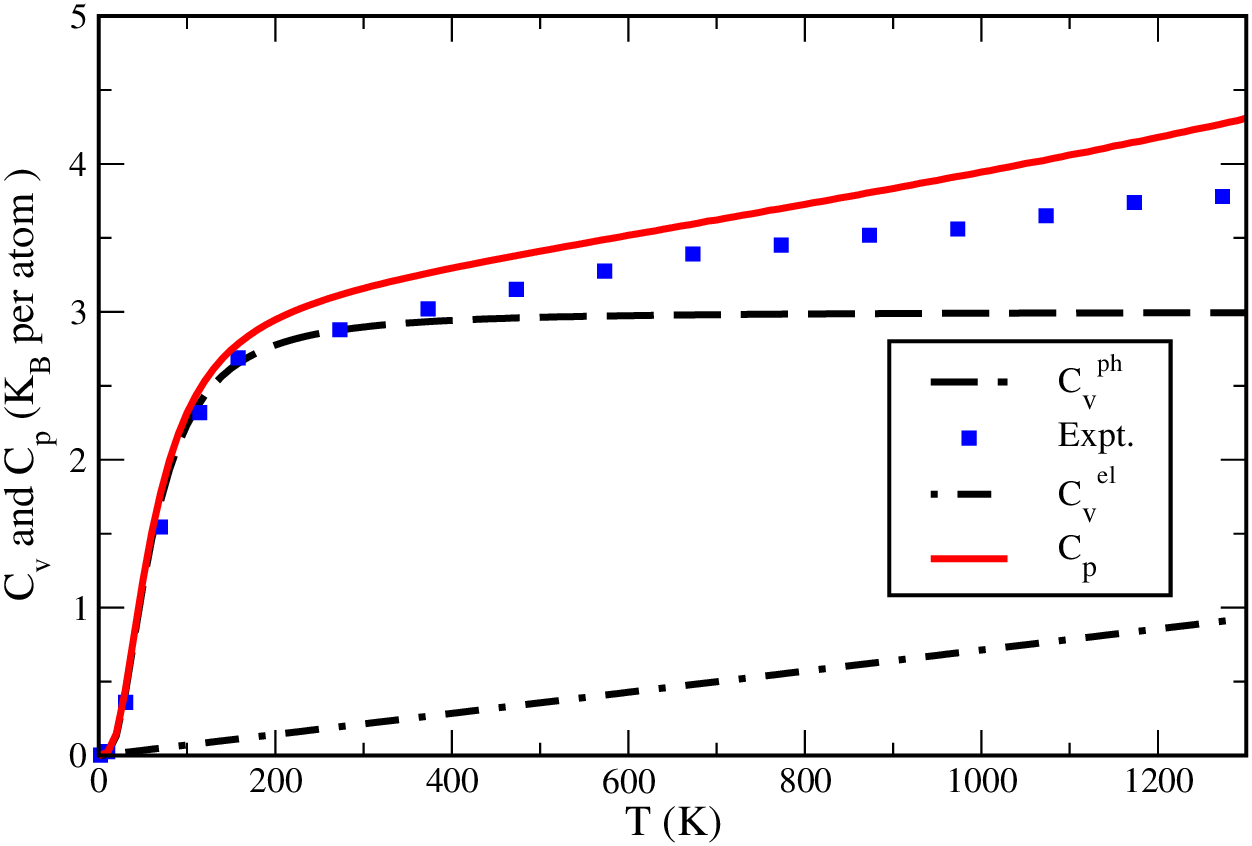}
\vspace{-0.05in}
\caption{Calculated temperature-dependence of heat capacity of Pd at constant volume 
$C_{V}^{ph}$, at constant pressure $C_{P}$ and the electronic contribution $C_{V}^{el}$. 
The square points are the experimental data for heat capacity at constant pressure from Ref.
[\onlinecite{Tou47}]. }
\label{cpPd}
\end{figure}

\begin{figure}
\renewcommand{\baselinestretch}{1}
\includegraphics[angle=0,width=7.6cm,clip]{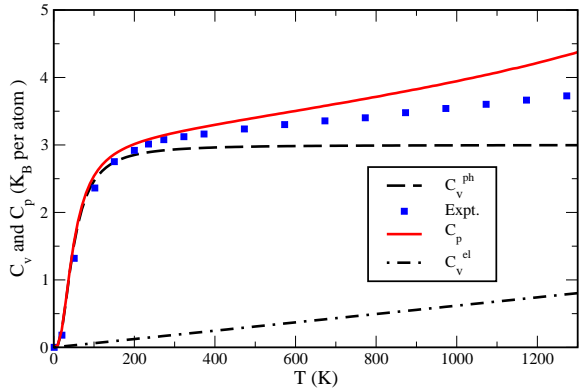}
\vspace{-0.05in}
\caption{Calculated temperature-dependence of heat capacity of Pt at constant volume
 $C_{V}^{ph}$, at constant pressure $C_{p}$ and the electronic contribution $C_{V}^{el}$.
  The square points are the experimental data for heat capacity at constant pressure 
  from Ref.  [\onlinecite{Tou47}]. }
\label{cpPt}
\end{figure}

\begin{figure}
\renewcommand{\baselinestretch}{1}
\includegraphics[angle=0,width=7.6cm,clip]{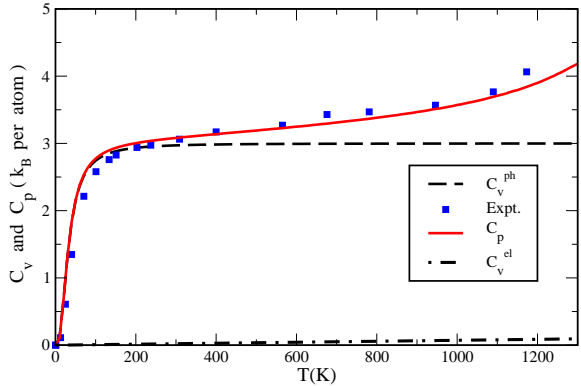}
\vspace{-0.05in}
\caption{Calculated temperature-dependence of heat capacity of Au at constant volume
 $C_{V}^{ph}$, at constant pressure $C_{P}$ and the electronic contribution $C_{V}^{el}$. 
 The square points are the experimental data for heat capacity at constant pressure from Ref.
 [\onlinecite{Tou47}]. }
\label{cpAu}
\end{figure}

Figs.(\ref {cpAg} - \ref {cpAu}) show the temperature-dependence of heat capacity at 
constant volume $C_{V}^{ph}$, at constant pressure $C_{P}$ and the electronic 
contribution $C_{V}^{el}$. The experimental results, which are for $C_{P}$, are 
also shown.  For Ni, Pd and Pt, where the Fermi level lies within (at the outer edge) 
the $d$-band, density of states $N(E_F)$ is large, resulting in a substantial 
contribution to $C_{V}$. In most cases the agreement with the experimental results 
is good, except for Ni which is ferromagnetic below the Curie temperature
$T_c$ of 627 K\cite{Kit53}. The discrepancy between the calculated and experimental 
results is largest at the Curie temperature and decreases steadily for temperatures  
both below and above $T_c$.  For both low and high temperatures, away from $T_c$, 
the agreement is very good. The discrepancy is understandable, as the present 
formulation of EAM does not distinguish between the magnetic
and nonmagnetic states. For Cu (Fig.  (\ref{cpCu})), the effects of anharmonicity 
can be seen in the dashed line, which represents the variation of $C_{P}$ with 
temperature obtained by Mei {\sl et al.}\cite{Mei1} using molecular dynamics 
simulation.  Note that despite poor agreement between the calculated and measured 
phonon spectra for Au, the calculated and measured values of $C_{P}$ agree very 
well over a large temperature range, possibly due to anharmonic effects excluded 
from the present calculation, which are, however, manifested in all measured 
properties.  Poor phonon frequencies and anharmonic effects compensate 
for each other, yielding excellent values for $C_{P}$.  Although the calculated 
phonon spectra for  both Au and Pt show poor agreement with measured values, 
measured $C_{P}$ for Au shows much better agreement with the calculated values 
than  for Pt. On the contrary, measured thermal expansion is much better 
reproduced for Pt than for Au by the calculation.  

\begin{figure}
\renewcommand{\baselinestretch}{1}
\includegraphics[angle=0,width=7.5cm,clip]{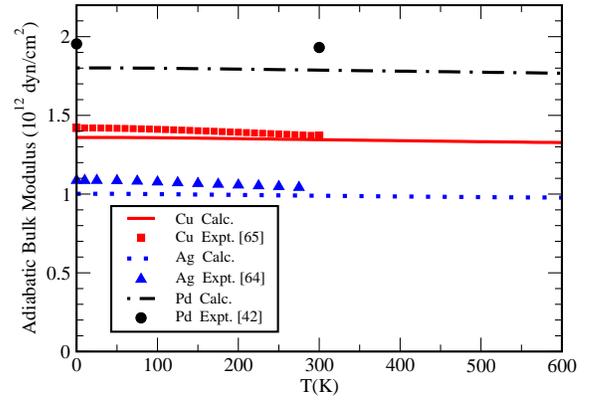}
\vspace{-0.05in}
\caption{Adiabatic bulk moduli $B(T)$ as a function of temperature for Ag, Cu and Pd.}
\label{BsAg}
\end{figure}

\begin{figure}
\renewcommand{\baselinestretch}{1}
\includegraphics[angle=0,width=7.5cm,clip]{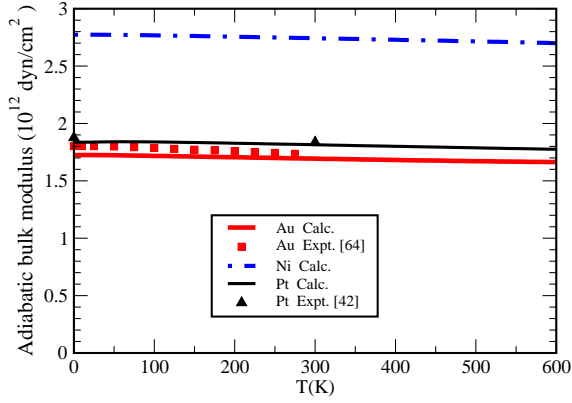}
\vspace{-0.05in}
\caption{Adiabatic bulk moduli $B(T)$ as a function of temperature for Au, Ni and Pt.}
\label{BsCu}
\end{figure}

The knowledge of $C_{P}$ and $C_{V}$ enables us to determine the adiabatic bulk 
modulus using the relation 
\begin{equation}
B_{S}(T)={C_{P}\over C_{V}}B(T). 
\end{equation}
Figs.(\ref{BsAg}-\ref{BsCu}) show the calculated results. Experimental results are 
indicated wherever available. Both the values and the trend, a slight decrease with 
increasing temperature, are reproduced quite well by the model.

\begin{figure}
\renewcommand{\baselinestretch}{1}
\includegraphics[angle=0,width=7.2cm,clip]{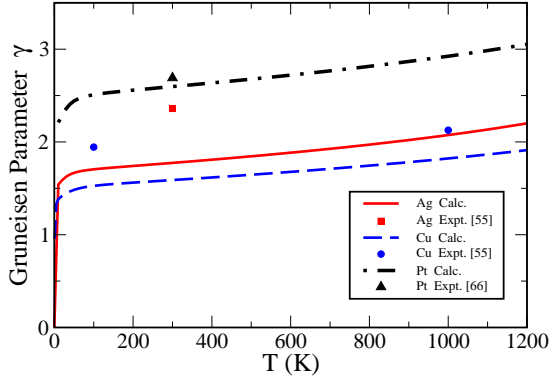}
\vspace{-0.05in}
\caption{Overall Gr\"uneisen Parameter $\gamma$ as a function of temperature for 
Ag, Cu and Pt.}
\label{gruAg}
\end{figure}

\begin{figure}
\renewcommand{\baselinestretch}{1}
\includegraphics[angle=0,width=7.2cm,clip]{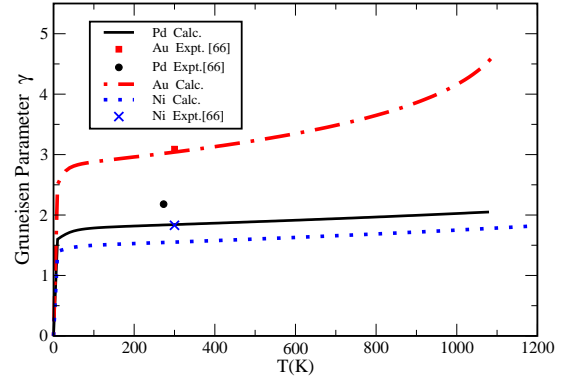}
\vspace{-0.05in}
\caption{Overall Gr\"uneisen Parameter $\gamma$ as a function of temperature for Au, Ni 
and Pd.}
\label{gruCu}
\end{figure}

An important thermodynamic property is the Gr\"uneisen  parameter, which is essentially 
a measure of the volume-dependence of the
phonon frequencies and can be related to thermal expansion\cite{ashcroft} of the solid.
 The mode-specific Gr\"uneisen Parameter corresponding to the $({\bf K},s)$ phonon mode 
 is given by
\begin{equation}
\gamma_{s}(\bf K)=-{V \over \omega_{s}(\bf K)} {\partial \omega_{s}(\bf K) \over \partial V}.
\end{equation}
We compute the overall Gr\"uneisen parameter $\gamma(T)$  by averaging over the individual
 Gr\"uneisen parameters $\gamma_{s}(\bf K)$ of all the modes with a weight of  
 $C_{V}({\bf K}s)$ from each mode$({\bf K}s)$\cite{ashcroft}, i.e.,
\begin{equation}
\gamma(T)={{\sum _{{\bf K}s} \gamma_{s}({\bf K}) C_{V}({\bf K}s) }\over {\sum_{{\bf K}s} 
C_{V}({\bf K}s)}}.
\label{gam}
\end{equation}
Figs. (\ref {gruAg} - \ref {gruCu}) show the calculated results for the variation 
of the overall Gr\"uneisen parameter $\gamma (T)$ as a function of temperature.

\begin{figure}
\renewcommand{\baselinestretch}{1}
\includegraphics[angle=0,width=7.2cm,clip]{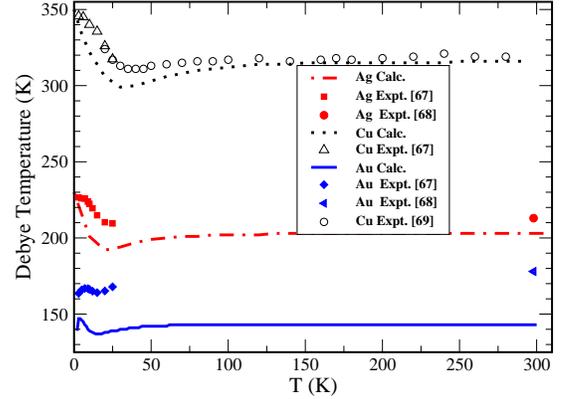}
\vspace{-0.05in}
\caption{Debye temperature $\Theta _D$ as a function of temperature $T$ for Cu, Ag 
and Au.}
\label{Debye1}
\end{figure}

\begin{figure}
\renewcommand{\baselinestretch}{1}
\includegraphics[angle=0,width=7.2cm,clip]{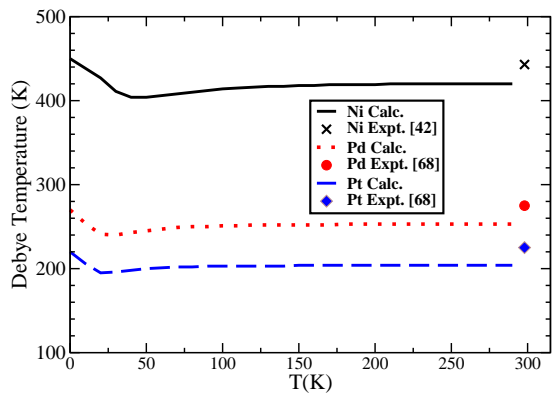}
\vspace{-0.05in}
\caption{Debye temperature $\Theta _D$ as a function of temperature $T$ for Ni, Pd 
and Pt.}
\label{Debye2}
\end{figure}

Finally, we study the Debye temperature and its temperature-dependence.
The temperature dependent Debye temperature $\Theta_{D}$ is obtained by numerically 
evaluating the integral 
\begin{equation}
C_{V}(T)=9k_{B}{\big (}{T\over \Theta_{D}}{\big )}^{3}\int_{0}^{X_{D}}{X^{4}{\rm exp}
(X)\over [{\rm exp}(X)-1]^{2}} dX 
\label{dCv}
\end{equation}
with 
\begin{equation}
X_{D}\equiv \Theta_{D}/T\; ,
\end{equation}
 and adjusting the value of $X_{D}$ so that the integral on the right hand side
 of Eq. (\ref{dCv}) matches the already computed value of $C_V(T)$ on the
 left hand side.
The results are plotted in Figs.(\ref{Debye1}) and (\ref{Debye2}). The calculated 
values are slightly smaller than the corresponding experimental values, reflecting 
the difference between the calculated and the measured values of $C_V$. However, 
the maximum difference is of the order of 10\%. For all metals there is an initial 
decrease in the Debye temperature as the temperature increases from zero, except 
for Au where there is an initial increase. It is interesting that the present EAM 
model is able to capture this trend.

\section {Summary of results and conclusions}
\label{conclusions}

This work presents the first complete study of the phonon spectra and all thermodynamic
properties of six fcc metals: Cu, Ag, Au, Ni, Pd and Pt, based on the EAM. The results
are obtianed using the
analytic embedding functions of Mei {\sl et al.}\cite{Mei1} in the quasiharmonic 
approximation.  The calculated phonon dispersion 
curves for Cu and Ag agree  well with the inelastic neutron diffraction results. The 
discrepancy between the calculated and the measured phonon frequencies increases with 
increasing phonon wave vector for all of the above metals.  However,  the relative error,  
at  the symmetry points $X$ and $L$, is not more than $ 5.0$\% for Ag and $7.0$\% 
for Cu.  Large differences between the calculated and the measured phonon frequencies are 
found for Ni, Pd, Pt and Au at high values of phonon frequencies, with the discrepancy for 
Au being the most drastic. However, the agreement or the lack thereof in the calculated 
phonon dispersion curves does not carry over to the thermodynamic properties in a 
proportionate way. For example, despite poor agreement for the phonon frequencies, 
the calculated and measured values of $C_P$ agree very well for Au over a wide 
temperature range. Although the calculated phonon spectra for  both Au and Pt show poor 
agreement with measured values, measured $C_{P}$ for Au shows much better agreement with 
the calculated values than  for Pt. On the contrary, measured thermal expansion
 is much better reproduced for Pt than for Au by the calculation.  

Both isothermal and adiabatic bulk moduli are very well represented over a wide 
temperature range for all the metals studied. This is perhaps not surprising, as the 
parameters in the model are determined largely by fitting to the
elastic constants, which usually have only very weak  temperature-dependence. In general, 
the co-efficient of thermal expansion, calculated in the quasiharmonic approximation, is 
underestimated for most of the metals beyond  room temperature. Calculated values of 
$C_P$ for Ag, Cu and Au agree very well with the measured values. For Ni the agreement
is good both below and above the Curie temperature, with the discrepancy increasing as 
the Curie temperature is approached from both above and below. This is understandable as 
the present model of EAM is not designed to capture the
physics of a ferromagnetic to paramagnetic transition. The Gr\"{u}neisen parameter and  
Debye temperature are underestimated by about 10\% for all the metals. However, the 
temperature variation of the Debye temperature, an initial
decrease as the temperature is raised from zero for all the metals except Au, is reproduced 
well by the model. For Au the experimental values of the Debye temperature show an initial 
increase. This feature is also captured by the model.

In summary, the thermodynamic properties of the six fcc metals: Cu, Ag, Au, Ni, Pd and Pt, 
and their temperature-dependence are reproduced reasonably well by the EAM model of Mei 
{\sl et al.}\cite{Mei1}. The coefficient of thermal expansion is underestimated above 
the room tempearture in the quasiharmonic approximation used in this
work. A better treatment of the anharmonicity in the potentials can improve the results, 
as evidenced by the molecular dynamics study of Mei {\sl et al.}\cite{Mei1}. There is 
room for much improvement in the phonon dispersion curves for Ni, Pd, Pt and Au. There 
is some evidence that this can be achieved by a change in the
fitting of the parameters. Preliminary calculations with small {\it ad hoc} changes in 
the parmeters representing the charge density in the model showed remarkable improvement 
in the phonon frequencies for Pd and Pt. This is an indication that a better representation 
of the charge desnity may lead to improvement for all of these metals.

\begin{center}
ACKNOWLEDGMENTS
\end{center}
Financial support for this work was provided by  Natural Sciences and Engineering 
Research Council of Canada.


\begin {thebibliography}{99}
\bibitem[$\dagger$] {byline} Present address: Department of Materials Science and engineering,
McMaster University, Hamilton, Ontario L8S 4L8, Canada.
\bibitem{pair-pot}{\it Interatomic  Potentials and the Simulation of Lattice Defects},
edited by P.C. Gehlen, J.R. Beeler, and R.I. Jaffee (Plenum, New York, 1972).
\bibitem{carlsson}A.E. Carlsson, C.D. Gelatt Jr., and H. Ehrenreich, Phil.
Mag. A {\bf 41}, 241 (1980); see also A.E. Carlsson, {\it Beyond Pair Potentials} in 
{\it Solid
State Physics}, Volume 43, edited by Henry Ehrenreich and David Turnbull, Academic
Press 1990, p.1. 
\bibitem{shukla} E. R. Cowley and R. C. Shukla, Phys. Rev. B {\bf 9}, 1261 (1974). 
\bibitem{Finnis26} M.W. Finnis and J.E. Sinclair,  Phil. Mag. {\bf A50}, 45 (1984).
\bibitem{Legrand} B. Legrand, Phil. Mag. B {\bf 49}, 171 (1984).
\bibitem{Rosato} V. Rosato, M. Guilope and B. Legrand, Phil. Mag. A {\bf 59}, 321 (1989),
and references therein.
\bibitem{Cleri} F. Cleri and V. Rosato, Phys. Rev. B {\bf 48}, 22 (1993).
\bibitem{Zhang} Z.J. Zhang, J. Phys: Condens. Matter {\bf 10}, L495 (1998).
\bibitem{Pettifor86} D.G. Pettifor, J. Phys. C: Solid State Phys. {\bf 19}, 285 (1986).
\bibitem{Pettifor87} D.G. Pettfor, Solid State Physics, Vol. 40, edited by H. Ehrenreich
and D. Turnbull (Academic Press, Inc., 1987), p. 43.
\bibitem{Ducastelle70} F. Ducastelle and F. Cyrot-Lackmann, J. Phys. Chem. Solids {\bf 31},
1295 (1970).
\bibitem{Ducastelle71} F. Ducastelle and F. Cyrot-Lackmann, J. Phys. Chem. Solids {\bf 32},
285 (1971).
\bibitem{Friedel} J. Friedel in {\it The Physics of Metals} (ed. J.M. Ziman), Cambridge Univ.
Press, 361 (1969)
\bibitem{gupta} R.P. Gupta, Phys. Rev. B {\bf 23}, 6265 (1981).
\bibitem{mukherjee} D. Tom\'{a}nek, S. Mukherjee, and K.H. Bennemann, Phys. Rev. B {\bf 28},
665 (1983).
\bibitem{sutton} A.P. Sutton, {\it Electronic Structure of Materials}, Clarendon Press, Oxford,
2004, Ch. 9.
\bibitem{pettifor} D.G. Pettifor, {\it Bonding and Structure of Molecules and Solids},
Oxford Science Publications, 2002, Ch. 7.
\bibitem{Daw2} M.S. Daw, and M.I. Baskes, Phys. Rev. Lett. {\bf 50} 1285 (1983).
\bibitem{Daw3} M.S. Daw, and M.I. Baskes, \PRB {\bf 29} 6443 (1984).
\bibitem{Manninen27} M. Manninen, \PRB {\bf 34} 8486 (1986).
\bibitem{Jacobsen28} K.W. Jacobsen, J.K. N$\phi$rskove and M.J. Puska, \PRB {\bf 35}, 7423 (1987).
\bibitem{Foiles6} S.M. Foiles, \prb {\bf 32} 3409 (1985).
\bibitem{Mei30} J. Mei and J.W. Davenport , \prb {\bf 42} 9682 (1991).
\bibitem{Mei-Al} J. Mei and J.W. Davenport, Phys. Rev. B {\bf 46} 21 (1992).
\bibitem{Foiles8} S.M. Foiles, M.I. Baskes and M.S. Daw, \prb {\bf 33} 7983 (1986).
\bibitem{Johnson9} R.A. Johnson,\prb {\bf 39}, 12554 (1989); \prb {\bf 41} 9717 (1990).
\bibitem{Daw5} M.S. Daw and R.D. Hatcher, Solid State Commun. {\bf 56}, 697 (1985).
\bibitem{Nelson17} J.S. Nelson, E.C.Sowa and M. S. Daw, Phys. Rev. Lett. {\bf 61} (1988).
\bibitem{Nelson18} J.S. Nelson, M.S. Daw, and E.C. Sowa, Phys. Rev. B {\bf 40}, 1465 (1989).
\bibitem{Daw11}  M.S. Daw and S.M. Foiles, J. Vac. Sci. Technol. A {\bf 4} 1412 (1986).
\bibitem{Daw11b} S.M. Foiles, Surf. Sci. {\bf 191}, L779 (1987).
\bibitem{Felter13} T.E. Felter, S.M. Foiles, M.S. Daw and R.H. Stulen, Surf. Sci. 
{\bf 171} L379 (1986).
\bibitem{Daw14}  M.S. Daw and S.M. Foiles, \PRB {\bf 35} 2128 (1987).
\bibitem{Johnson4} R.A. Johnson, \prb {\bf 37} 3924 (1988).
\bibitem{Daw15} M.S. Daw and S.M. Foiles, Phys. Rev. Lett. {\bf 59}, 2756 (1987).
\bibitem{Foiles16} S.M. Foiles, Surf. Sci. {\bf 191}, 329 (1987).
\bibitem{Bas10} M.I. Baskes, \PRB {\bf 46}, 2727 (1992).
\bibitem{foiles-adams} S.M. Foiles and J.B. Adams, Phys. Rev. B {\bf 40}, 5909 (1989).
\bibitem{OhJohnson}D.J. Oh, and R.A. Johnson, J. Mater. Res. {\bf 3}, 471 (1988).
\bibitem{Mei1} J. Mei , J.W. Davenport and G.W. Fernando, \prb {\bf 43}, 4653 (1991).
\bibitem{kuiying} C. Kuiying, L. Hongbo, L. Xiaoping, H. Qiyong, and
H. Zhuangqi, J. Phys: Condens. Matter {\bf 7} 2379 (1995).
\bibitem{Kit53} C. Kittel, {\sl  Introduction to Solid State Physics} (John Wiley and Sons,
 Inc., New York, 1996).
\bibitem{EC34} E.C. Svensson, B.N. Brockhouse and J.M. Rowe, \prb {\bf 155}, 619 (1967).
\bibitem{nielsson} G. Nilsson and S. Ronaldson, Phys. Rev. B {\bf 7}, 2393 (1973).
\bibitem{clementi} E. Clementi and C. Roetti, {\sl Atomic Data and Nuclear Tables}, 
Vol. 14, 177 (1974).
\bibitem{Mclean} A.D. McLean and R.S. McLean, {\sl Atomic Data and Nuclear Tables}, 
Vol. 26, 197 (1981).
\bibitem{savrasov} see, for example,  the full potential linear muffin-tin 
(FP-LMTO) results for Cu and Pd by
S.Y. Savrasov and D.Y. Savrasov, Phys. Rev. B {\bf 54}, 16487 (1996).
\bibitem{papa2} see Y. Mishin, M.J. Mehl, D.A. Papaconstantopoulos,
A.F. Voter and J.D. Kress, Phys. Rev. B {\bf 63}, 224106 (2001). For the test of 
Cu studied by the
authors, {\it ab initio} tight binding results for high frequency phonons are 
superior to those given by the two EAM schemes studied.
\bibitem{sobhana} see, for example, the pseudopotential-based results for Cu by 
S. Narasimhan and S. de Gironcoli, Phys. Rev. B {\bf 65}, 064302 (2002). 
The results are somewhat inferior to those based on the
FP-LMTO method of Ref. [\onlinecite{savrasov}] or the {\it ab initio} 
tight-binding results of Ref. [\onlinecite{papa2}], and also seem to depend on 
the scheme, LDA (local density approximation) vs. GGA (generalized gradient approximation),
used to treat the exchange-correlation potential.
\bibitem{WA33} W.A. Kamitakahara and B.N. Brockhouse, Phys. Lett. {\bf 29A}, 639 (1969).
\bibitem{Lynn38}  J.W. Lynn, H.G. Smith and R.M. Nicklow, \prb {\bf 8}, 3493 (1973).
\bibitem{Birgr37} R.J. Birgeneau, J. Cordes, G. Dolling and A.D.B.woods, 
\prb {\bf 136}, A1359 (1964).
\bibitem{Miller35} A.P. Miller and B.N. Brockhouse, Can. J. Phys. {\bf 49}, 704 (1971).
\bibitem{Dutton36} D.H. Dutton and B.N. Brockhouse, Can. J. Phys. {\bf 50}, 2915 (1972). 
\bibitem{Mac44} R.A. MacDonald and W.M. MacDonald, \prb {\bf 24}, 1715 (1981).
\bibitem{shukla-mac} R.C. Shukla and R.A. MacDonald, High Temp.-High Press. 
{\bf 12}, 291 (1980).
\bibitem{Tou43} Y.S. Touloukian, R.K. Kirby, R.E. Taylor and P.D. Desai, 
{\sl Thermophysical Properties of Matter, The TPRC Data Series, Thermal Expansion, 
Metallic Elements and alloys,} (Plenum Data Company,   New York, 1977).  
\bibitem{ole} O.K. Andersen
Phys. Rev. B {\bf 8}, 3060 (1975);
 O.K. Andersen, O. Jepsen, and  D. Gl\"{o}tzel, 1985, in {\em
 Highlights of Condensed Matter Theory}, edited by F. Bassani, F. Fumi,
 and M.P. Tosi, North-Holland, Amsterdam, 1985, pp. 59-176;
 O.K. Andersen, O. Jepsen, M. Sob in {\em Electronic structure
and its applications}, edited by M. Yossouff, Lecture Notes in Physics, v.283
(Springer, Berlin,1987) p.1-57; see also  http://www.fkf.mpg.de/andersen/.
 \bibitem{papa-MJW} see for example "{\it Calculated electronic properties of metals}"
 by V.L. Moruzzi, J.F. Janak and A.R. Williams, Pergamon, New York 1978;
 "{\it Handbook of the band structure of elemental solids}" by D.A.
 Papaconstantopoulos, Plenum, New york, 1986.
\bibitem{grimvall} G\"{o}ran Grimvall, {\it The Electron-phonon Interaction in Metals},
North-Holland, Amsterdam, Ch. 6 (1981).
\bibitem{prange} R.E. Prange and L.P. Kadanoff, Phys. Rev. {\bf 134}, A566 (1964).
\bibitem{Wal53} D.C. Wallace, {\sl Thermodynamics of Crystals} (John Wiley and Sons, Inc.,
Toronto, 1972).
\bibitem{Tou47} Y.S. Touloukian and E.H. Buyco, {\sl  Thermophysical Properties of Matter, 
The TPRC Data Series, Specific Heat, Metallic Elements and alloys, Vol.4} 
(Plenum Data Company, New York, 1970).
\bibitem{Nei55} J.R. Neighbours and G.A. Alers, Phys.Rev. {\bf 111}, 707 (1958).
\bibitem{Over} W.C. Overton,Jr. and J. Gaffney, Phys. Rev. {\bf 98}, 969 (1955).
\bibitem{Pan} C.V. Pandya, P.R. Vyas, T.C. Pandya and V.B. Gohel, Bull. Mater. Sci {\bf 25},
 63 (2002).
\bibitem{Mar50} D.L. Martin, Phys. Rev. {\bf 141}, 576 (1966).
\bibitem{Com51} W.D. Compton, K.A. Gschneidner, M.T. Hutchings, H. Rabin and M.P.Tosi,
 {\sl Solid States Physics, Advances in Research and Applications, Vol.16} (Academic Press,
 New York and London, 1964).
\bibitem{Mart52} D.L. Martin, Can. J. Phys. {\bf 38}, 2049 (1960).
\bibitem{ashcroft} N.W. Ashcroft and N.D. Mermin, {\sl Solid State Physics} (Saunders
College, Philadelphia, 1976), p. 493.
\end{thebibliography}
\end{document}